\documentclass[useAMS,usenatbib]{mn2e}

\usepackage{graphicx}
\usepackage{amssymb}
\usepackage{txfonts}
\usepackage{biblio}
\bibpunct{(}{)}{,}{a}{}{,}

\newcommand{\VHf}{Voigt-Hjerting function}

\newcommand{\ie}{\textit{i.e.}}

\newcommand{\eg}{\textit{e.g.}}

\newcommand{\hi}{H{\sc i}}

\newcommand{\nhi}{\mathrm{N_{\,H I}}}

\newcommand{\los}{line-of-sight}

\newcommand{\loss}{lines-of-sight}

\newcommand{\cmsq}{\, \mathrm{cm}^{\,-2}}

\newcommand{\nm}{\, \mathrm{nm}}

\newcommand{\kms}{\, \mathrm{km \, s^{\,-1}}}

\newcommand{\diff}{\, \mathrm{d} \,}

\newcommand{\vel}{\mathrm{v}}

\newcommand{\lya}{Ly$\alpha$}

\newcommand{\lyb}{Ly$\beta$}

\newcommand{\da}{D_{\,\mathrm A}}

\newcommand{\zem}{z_{\,em}}

\title[]{Stochastic Absorption of the Light of Background Sources due to Intergalactic Neutral Hydrogen\\
{\Large I. Testing different line-number evolution models via the cosmic flux decrement}}

\author[T. Tepper-Garc\'\i a and U. Fritze]{Thorsten Tepper-Garc\'\i a$^{1}$\thanks{Currently at: Institut f\"ur Physik, Universit\"at Potsdam, Am Neuen Palais 10, D-14469 Potsdam; E-mail:
tepper@astro.physik.uni-potsdam.de} and Uta Fritze$^{2}$\\
$^{1}$Institut f\"ur Astrophysik, Georg-August Universit\"at, Friedrich-Hund-Platz 1, D-37077 G\"ottingen\\
$^{2}$Centre for Astrophysics Research, University of Hertfordshire, College Lane, Hatfield, AL10 9AB, UK
}
\begin{document}

\date{Accepted ---. Received ---; in original form ---.}

\pagerange{\pageref{firstpage}--\pageref{lastpage}}
	
\pubyear{2005}

\maketitle
	
\label{firstpage}
	
\begin{abstract}
		We test the accuracy of different models of the attenuation of light due to resonant scattering by intergalactic neutral hydrogen by comparing their predictions of the evolution of the mean cosmic flux decrement, $\da$, to measurements of this quantity based on observations. To this end, we use data available in the literature and our own measurements of the cosmic flux decrement for 25 quasars in the redshift range $2.71 < \zem < 5.41$ taken from the SDSS Data Release 5. In order to perform the measurements of $\da$, we fit a power-law to the continuum redward of the \lya{} emission line, and extrapolate this fit to region blueward of it, where the flux is severely affected by absorption due to intervening \hi{} absorbers.

	We compute, using numerical simulations, the redshift evolution of $\da$ accounting for the presence of \lya{} Forest absorbers and Lyman limit systems randomly distributed along the \los{}, and compute its intrinsic scatter at the 1-, 2-, and $3 \, \sigma$ level due to fluctuations in the absorber properties (column density, Doppler parameter, redshift) along different \loss{}. The numerical simulations consist of Monte Carlo realizations of distributions of the absorber properties constrained from observations.

	The results from the models considered here confirm our theoretical expectation that the distribution of $\da$ at any given redshift be well described by a lognormal distribution function. This implies that the effective optical depth, usually defined as the negative logarithm of the average flux, 1 - $\da$, is very accurately Gaussian distributed, in contrast to previous studies. This result is independent to the form of the input distribution functions, and rather insensitive to the presence of high-column density absorbers, such as the Lyman limit systems.

	By comparing our and previous measurements of $\da$ to the outcomes of our simulations, we find an excellent agreement between the observations and the evolution of the \emph{mean} $\da$ as predicted by one of the models considered in this work. The \emph{observed scatter} in $\da$ at each redshift, however, cannot be recovered from our simulations. Even though there is evidence for the fact that the lack of agreement between models and observations comes from the combination of heterogeneous measurement sets obtained by different methods, the failure of the models to accurately account for the absorption by intergalactic \lya{} absorbing systems and its variation along different \loss{} cannot be completely ruled out.
\end{abstract}
	

%---------------------------------------------------------------------------------------------------------------
\section{introduction} \label{sec:int}

	Since the introduction of the Gunn-Peterson (GP) test by \citet*{gun65a}, a detailed knowledge about the physical state of the intergalactic medium (IGM) has been gained from the study of the absorption features identified in the spectra of quasi-stellar objects (QSOs) at restframe wavelengths $\lambda \leq 121.5 \nm$, which are now known to be mainly due to resonant scattering by intergalactic neutral hydrogen randomly distributed along the \los{}, as first proposed by \citet{lyn71a}. For instance, the null result in the search for a GP trough has been used to rule out the existence of a hot intercloud medium (ICM) \citep*{ste87b,gia92a,gia94a}, which was thought to confine by pressure the \lya{} clouds \citep*{sar80a,ost83a}. As a result of detailed analyses of the line statistics of the absorbing material, a wealth of information on its clustering \citep*[see \eg{}][]{ost88a}, in particular the existence of voids \citep*[\eg{}][]{pie88a,dun89a,dob91a}, and the evolution of its number densities, column densities, and Doppler parameter with redshift \citep[\eg{}][]{kim97a} has accumulated over the past years. These results, in combination with the use of state-of-the-art numerical simulations of structure formation based on the currently accepted paradigm of the concordance cosmology \citep{spr05a}, show that the features seen in absorption against bright background sources arise when the \los{} intersects the structures that naturally emerge and evolve with time under the influence of gravitational attraction. Different types of structures such as the filaments present in the \emph{cosmic web}, galactic haloes, and even the discs of primeval galaxies, give rise to distinct absorption features attributed to entities known historically as \lya{} Forest clouds, Lyman limit systems (LLSs), and damped \lya{} absorbers (DLAs) \citep*[see \eg{}][for excellent reviews, respectively]{rau98a,wol05a}. Nevertheless, the relation between the observed absorption features and the objects causing them, in particular the correlation between observed damped absorption lines, metal lines (\eg{} Mg {\sc ii}, O {\sc vi}) and galaxies--the so-called Absorber-galaxy Connection--is still a matter of debate \citep[see \eg{}][Part 1]{05a}. As a consequence of numerous efforts over many years, we now have a better understanding of the origins of the different absorption features observed in QSO spectra. In particular, the notion of discrete, intervening \hi{} absorbing systems randomly distributed along the \los{} has been embedded into the more general picture of an evolving \emph{continuous} intergalactic medium with a \hi{} density field that varies in space and time, with its evolution driven mainly by the Hubble expansion, the radiation field of ionising UV sources, and the collapse of structures due to gravity.

%------------------------------------------------------------------------------------------------------------------------------------------
\subsection{Methods and Input Distributions: A Brief Review} \label{sec:rev}

	Over decades many people have been working hard towards inferring the physical properties of the intergalactic medium such as its chemical content, density, temperature, etc. \citep*[see \eg{}][]{kim02b} by measuring the type of transition, strength, number density, and profiles of absorption lines imprinted in the spectra of QSOs and, more recently, of Gamma-Ray bursts (GRBs) \citep*[\eg{}][]{lam00a}. There has also been a great effort to quantify the effect of the absorption due to intergalactic neutral hydrogen on the photometric properties of background sources. As a matter of fact, several models have been developed in order to account for this so-called intergalactic attenuation, with different approaches and purposes. \citet*{mol90a} used Monte Carlo simulations to estimate the amount of absorption at wavelengths shorter than the redshifted He{\sc ii} $\lambda \, 30.4 \nm$ line, in order to test the feasibility of the equivalent of the Gunn-Peterson test for intergalactic helium. They found that the absorption as a function of wavelength, averaged over many \loss{}, should display together with a characteristic stair-case profile due to the cumulative absorption at the \hi{} resonant wavelengths, an additional characteristic valley-shaped feature (the ''Lyman-valley'') due to the cumulative effect of the photoionisation of \hi{} by photons with energies $E_{\,\gamma} \geq h \, c / \lambda_{\,L}$, where $\lambda_{\,L} = 91.2 \nm$.
	Later on, in a seminal paper \citet{mad95a} developed an analytical method to quantify the opacity due to intergalactic \hi{} as a function of redshift, and its effect on the colors of high-redshift galaxies. The underlying assumption of this model is that the observed flux of a source at redshift $z$ is given by the product of the \emph{intrinsic} flux and a transmission factor that accounts for the \emph{mean} absorption as a function of wavelength given in the form $\exp(-\tau_{\,\mathrm{eff}}) \equiv \langle \exp(-\tau) \rangle$ \citep[see][their equation 3]{mad95a}, where the brackets denote the average over an ensemble of random \loss{}. The most common application of this model consists in correcting the flux of a synthetic (galaxy-, QSO-) spectrum for intergalactic absorption. This correction is of particular importance at high redshift, where intergalactic \hi{} severely absorbs the light of a background object at restframe wavelengths shorter than $121.6 \nm$, leading to a substantial reddening of its colour \citep[see \eg{}][]{bic04a}. As the numerous references in the literature attest, the \citeauthor{mad95a} model has become the most widely used attenuation model. However, in a later work \citet[][from now on BCG99]{ber99a} argued that it is not possible to estimate the mean change in the magnitude of a source at a given redshift due to absorption by intergalactic \hi{} along the \los{} by multiplying the mean transmission curve of \citeauthor{mad95a}'s model with the spectrum of the source and integrating over the corresponding passband, mainly because of the existence of color terms. They suggested that the correct way of accounting for the mean effect of \hi{} absorption on the spectrum of a background source and on its photometric properties is to model first the absorption along many random \loss{}, compute the desired photometric quantities for each one of them, and \emph{then} compute the average over the ensemble of \loss{}. In other words, they argue that the processes of averaging over many random \loss{} and measuring photometric quantities are non-commutative. Indeed, they showed using a Monte Carlo technique that the average magnitudes computed following their approach substantially differ from those computed using \citeauthor{mad95a}'s model, even when using the same input distributions for the number of absorbers, their column densities and Doppler parameters. The approach proposed by \defcitealias{ber99a}{BCG99}\citetalias{ber99a} effectively mimics the measurement process that would take place if one would determine \eg{} the mean observed brightness of a collection of galaxies with different absorber populations along their particular \loss{}, but otherwise identical in their intrinsic properties (spectrum, morphology, etc.), and is hence physically meaningful.  It turns out that features such as the characteristic stair-case profile and the Lyman-valley cannot possibly be observed in a \emph{single} spectrum, since they arise only by averaging over sufficient numbers of \loss{}, a process that has no physical meaning.

	In a more recent paper, \citet{mei06a} developed a method to compute the opacity due to intergalactic \hi{} by using hydrodynamical simulations of structure formation in the framework of the concordance $\Lambda$CDM cosmology performed by \citet*{mei04a}. Applying their model to compute broad-band magnitudes for different types of objects (\eg{} starburst galaxies, QSOs of Type I and II), \citet{mei06a} reports differences of 0.5 -- 1.0 mag with respect to \citet{mad95a}'s model. Despite the different results obtained, this model is similar to \citet{mad95a}'s model in the sense that it implicitly assumes that the mean opacity of the IGM along a random \los{} due to the presence of \hi{} can be accounted for by multiplying a given input spectrum with a mean attenuation curve of the form $\exp(-\tau_{\,\mathrm{eff}})$ and integrating over the corresponding filter function \citep[see][their equation 8]{mei06a}.

	Following \citetalias{ber99a}, we state that
	\begin{equation}
		\int_{\,0}^{\,\infty} \, f_{\,\lambda} \cdot \langle \exp(-\tau) \rangle \cdot T(\lambda) \diff 			\lambda \neq \left \langle \int_{\,0}^{\,\infty} \, f_{\,\lambda} \cdot \exp(-\tau) \cdot T \, 			(\lambda) \diff \lambda \right \rangle \, ,
	\end{equation}
	where $f_{\,\lambda}$ is the intrinsic flux, $T(\lambda)$ is the filter transmission function, and the brackets denote the average over all \loss{}. We consider that the operation implied by the right-hand side of this expression is the correct way of estimating mean magnitudes of background objects including the effect of the absorption due to intergalactic \hi{}. This approach is of course not restricted to the computation of mean magnitudes and colors, and can be applied to the estimate of the mean of any photometric quantity. Furthermore, it is also possible to determine not only the mean, but in principle any desired confidence interval around the mean, \eg{} $\pm \, \sigma$ range, via the computation of quantiles (see Section \ref{sec:res}).

	It should be clear that not only the method, but also the input physics is an (even more) crucial ingredient of a particular model that accounts for the intergalactic attenuation, as already shown by \citetalias{ber99a}. It is, however, not trivial to test whether using a particular method and a set of input distributions accurately describes the observed effect of the absorption by intergalactic \hi{} on the spectra of background sources. For example, the evolutionary synthesis models of \citet{bic04a} that include the correction for intergalactic absorption based on \citeauthor{mad95a}'s model match quite well the observations of galaxies in the Hubble Deep Field \citep[][their Figure 12]{bic04a}, since the magnitude differences reported by \citetalias{ber99a} (their Figure 7) for one of their models with respect to \citeauthor{mad95a}'s model are in this case of the order of the scatter of the observations around the predicted colors. In other words, even though \citeauthor{mad95a}'s and \citetalias{ber99a}'s approaches are fundamentally different, it is difficult to test the accuracy of their predictions on the basis of a comparison to \eg{} observed galaxy colors. A quantity that is more sensitive to the absorption due to intergalactic \hi{} is the mean cosmic flux decrement $\da$ (cf. Section \ref{sec:cfd}). The reason for this is that the restframe wavelength range over which this quantity is measured is typically $10 \nm$ wide, and is hence narrower than typical broadband filters. We thus consider as a primer test that any model that accounts for the absorption due to intergalactic \hi{} should reproduce first of all the observations of this quantity. Hence, through the comparison of their respective predictions to measurements of $\da$ it should be possible to discriminate between different models which may or may not be appropriate to account for the effect of the intergalactic attenuation on the spectra of background sources. We select for this comparison two of several models introduced by \citetalias{ber99a} to compute the magnitude changes of high-redshift galaxies due to intergalactic absorption, and which are based, respectively, on the input distribution functions of \citet{kim97a}, and the input physics of the \citet*{mad95a} approach. The latter is chosen since, as already mentioned, it is the model most widely used in the literature; the former constitutes, to the best of our knowledge, the most complete set of input distribution functions for the evolution of the \lya{} absorbers to date, later expanded but not significantly changed by \citet{kim02a}.\\

	To sum up: The main goal of this work is to model and analyse the redshift evolution of $\da$ for different evolution scenarios of the intergalactic neutral hydrogen, conveniently parametrized by input distribution functions of the form of equation (\ref{eq:ddf}, Section \ref{sec:cfd}). Hence, the models we consider here differ only by the set of input distribution functions used, but they all equal in method, \ie{} all of them take advantage of the Monte Carlo technique. We judge the goodness of a particular model by its power to reproduce the observations of $\da$ in a wide redshift range, and analyse to which extent the theoretical expectations about the properties of $\da$ are recovered from the measurements.

	This work is organised as follows: In Section \ref{sec:cfd} we briefly recall the concept of the cosmic flux decrement and discuss some issues related to its measurement. In Section \ref{sec:mod} we present two different types of models for the intergalactic attenuation, which we use to model the redshift evolution of the cosmic flux decrement. In Section \ref{sec:mda}, we describe our measurements of this quantity for a sample of SDSS QSO spectra. Finally, we compare these and previous measurements to the outcomes of each model, and discuss the results of this comparison as well as some other implications of the models for the evolution of $\da$ in Section \ref{sec:res}.

%--------------------------------------------------------------------------------------------------------------
\section{The Cosmic Flux Decrement Revisited} \label{sec:cfd}

	Before high-resolution (\ie{} $\Delta\lambda \lesssim 1 \nm$), high S/N observation became feasible, the basic spectroscopic technique used to analyse the effect of the absorption due to intergalactic neutral hydrogen on the spectra of background sources was to measure the mean depression of the observed flux relative to the unabsorbed flux --or emission continuum--, a quantity which became to be known as cosmic flux decrement. This quantity, first introduced by \citet*{oke82a}, can be defined as a function of redshift by

	\begin{equation} \label{eq:cfd}
		\da \, (z) \equiv \frac{1}{\Delta \lambda} \int_{\,\lambda_{\,1}\cdot(1+z)}^{\lambda_{\,2}\cdot(1+z)} \left(1 - \frac{ f_{\,obs}(\lambda)}{f_{\,c}(\lambda)} \right )\mathrm{d} \lambda \, ,
	\end{equation}
	where $f_{\,c}$ and $f_{\,obs}$ are the continuum and the observed fluxes, respectively, and \mbox{$\Delta \lambda \equiv (1+z)\cdot(\lambda_{\,2} - \lambda_{\,1})$}. Formally, the integral is computed in the restframe wavelength range \mbox{$[102.5, \, 121.6] \nm$}, \ie{} between the \lyb{} and the \lya{} emission lines. However, the actual estimate of $\da$ is usually performed between the restframe wavelengths \mbox{$\lambda = 105 \nm$} and \mbox{$\lambda = 117 \nm$ }--or in an even narrower wavelength interval-- in order to avoid contamination by the emission wings of the \lyb{} + O{\sc vi} and \lya{} lines, respectively.

	Since $\da$ is extremely sensitive to $f_{\,c}$, as can be easily seen from the definition (\ref{eq:cfd}), an accurate measurement of this quantity demands a reliable estimate of the underlying continuum. Unfortunately, there is no consensus of what the best method to estimate the continuum may be. A popular choice, mainly because of the presence of emission lines redward of the \lya{} forest region, consists in fitting a local continuum, most commonly using cubic splines \citep*[see \eg{}][]{lu96a} or b-spline functions \citep*[see \eg{}][]{kir03a,tyt04a,tyt04b}, searching for regions apparently free of absorption blueward of the \lya{} emission line. Other authors prefer to fit a continuum in the region redward of the red wing of the \lya{} emission line, and extrapolate it to the region blueward of it \citep*{ste87a,sch89a,cri93a}. A widely adopted form for the fitted continuum in this case is a power-law with spectral index $\alpha_{\,\nu}$ which for wavelengths $\lambda > 121.6 \nm$ takes on values in the range \mbox{$[0.28,\,0.99]$} \citep*[][and references therein]{ste87a,van01a}.  The latter method may tend to place the intrinsic continuum level higher than it actually is, thus overestimating the measured values of $\da$ \citep[cf. Sec. \ref{sub:con}; see also][and references therein]{tyt04a}; for the former method the opposite is true, in general \citep*[see \eg{}][]{fau07a}. For either method, there is an uncertainty in the estimate of the continuum, and this is the main drawback of the mean flux depression as a technique to estimate the mean absorption due to neutral hydrogen present in the intergalactic medium (IGM). In an attempt to overcome this problem, \citet*{zuo93c} used the idea that the cosmic flux decrement effectively measures the \emph{total} equivalent width of all \lya{} absorption lines in the chosen wavelength range --if corrected for the contribution of metal absorption lines-- to measure this quantity by simply adding up the equivalent widths of lines identified as \lya{} absorption lines. Of course, the reliability of this technique highly depends on the correct identification of lines, and misidentification may lead to larger uncertainties than those associated to the continuum estimate.

	It turns out that reliable measurements of $\da$ are very useful to constrain estimates of fundamental cosmological parameters such as the mean baryon density $\Omega_{\,b}$, the UV background intensity \citep*{rau97b}, the normalization of the power spectrum $\sigma_{\,8}$, the vacuum-energy density $\Omega_{\,\Lambda}$, and the Hubble parameter $H_{\,0}$  \citep{tyt04a}. It is hence of great interest to contrast observations to theoretical models of the evolution of $\da$, in which the bias due to the uncertainty in the estimate of the continuum is absent. This has been previously done by different workers \citep[see \eg{}][]{gia90a,cri93a,mad95a}, usually obtaining a good agreement with observations. However, there is still a scatter in the observations of this quantity for which it has not been accounted yet in any modelling so far. It is possible that this scatter may well be due to the difference in the methods used by each group to measure $\da$ and to the different redshift ranges probed. Hence, by taking advantage of the Monte Carlo technique, we assess to which extent the \emph{observed} scatter can be ascribed to the \emph{intrinsic} scatter in $\da$ due to fluctuations in the properties (number density, column density, Doppler parameter) of the \hi{} absorbers along different \loss{} (see Section \ref{sec:sca}).

	From the theoretical point of view, assuming that the restframe equivalent width of the absorbers does not evolve with redshift, and that the number density of the absorbing systems evolves like $ \propto (1+z)^{\,\gamma}$, it is expected that $\da$ should evolve with $z$ like
	\begin{equation} \label{eq:pwl}
		D_{\,\mathrm A}(z) \propto (1+z)^{\,1 + \gamma}
	\end{equation}
	where the extra factor $(1+z)$ comes from the scaling of the equivalent width, as pointed out by \citet*{jen91a}. Indeed, it has been found empirically that the redshift evolution of $\da$ can be described by a power law \mbox{$\da\,(z) = A \cdot (1 + z)^{\,1+\gamma}$} with $A = 6.2 \cdot 10^{\,-3}$ and $\gamma = 1.75$ \citep{kir05a}, even though other functional forms, \eg{} an exponential of the form \mbox{$\da\,(z) =  \da^{\,0} \cdot \mathrm{e}^{\,\alpha \cdot (1+z)}$} with $\da^{\,0} = 0.01$ and $\alpha = 0.75$ \citep{zha97a} also match well the observations. More recently, \citet{kir07a} showed that the observed evolution of $\da$ with redshift in the range $0 < z < 3.2$ is well described by a broken power-law, even though the significance of the fit is low. In any case, expressions like these are only valid up to a given redshift, since they diverge for $z \to \infty$, whilst $\da$ converges asymptotically to 1 in this limit or, more precisely, when z approaches the redshift $z_{\,reion}$ at which reionisation sets on. As will be shown later, the redshift evolution of $\da$ predicted by our simulations does satisfy this asymptotic behaviour (cf. Section \ref{sec:res}). Furthermore, even if the power-law form for the evolution of $\da$ holds, the index $\gamma$ in equation (\ref{eq:pwl}) should be replaced by $\overline{\gamma}$, where the latter index accounts for the averaged evolution of absorbers of different column densities, which evolve all at different rates. Conversely, estimates of a single $\gamma$ from $\da$ measurements assuming a power-law of the form of equation (\ref{eq:pwl}), as done by \citet*{obr88a}, may give a hint on the population of absorbers {\it dominating} the behaviour of $\da$, comparing the estimated $\gamma$ with the power-law index of the different populations. For instance, according to one of the models considered here, the redshift evolution of $\da$ will be shown to be dominated by absorbers with column densities $\nhi < 10^{\,17} \cmsq$.\\

	A compilation of $\da$ measurements, accumulated in the literature over the past twenty years approximately, and which includes our own measurements that extend the redshift range to $\zem = 5.41$, is shown in Figure \ref{fig:daz} (cf. Section \ref{sec:mda} and cited references for details on the measurements in this figure).

%--------------------------------------------------------------------------------------------------------------
\section{Modelling The Intergalactic Attenuation} \label{sec:mod}

	Since the observation of individual sources (galaxies, QSOs, GRBs) necessarily implies observations along different \loss{}, it is expected that the stochastic nature of the distribution of the \lya{} absorbers, especially of those with the highest column densities, causes a scatter in the observed absorption, even for sources with identical intrinsic spectra. Hence, depending on the absorption along a particular \los{}, one would expect different observed values for each measurement of any photometric quantity, for example, the cosmic flux decrement $\da$. Performing enough measurements of such a quantity for sources with --ideally-- identical SEDs at a fixed redshift, one could in principle estimate its mean and its scatter due to stochastic effects in the absorption by neutral hydrogen in the IGM.

	The numerical realisation of this thought experiment is best achieved through Monte Carlo simulations. Following \eg{} \cite*{mol90a,gia90a,cri93a}, and \citetalias{ber99a}, we generate thousands ($4 \cdot 10^{\,3}$) of \loss{}  each with a random population of \hi{} absorbers, and compute the absorption along each of them for a given input spectrum at a fixed redshift. The population of each \los{} consists of a random number $N_{abs}$ of absorbing systems, each of them characterized by three parameters: its redshift $z_{\,abs}$, its column density $\nhi$, and its Doppler parameter \mbox{$b \equiv \sqrt{2kT/m_H}$}, where $k$ is the Boltzmann constant, $T$ is the kinetic temperature of the gas and $m_H$ is the mass of the hydrogen atom.

	The redshift and column density characterising each absorber are drawn from a distribution of the form
	\begin{equation} \label{eq:ddf}
 		f\,(\nhi,z) = \mathcal{N}_{\,0} \cdot \left (1 +
		z \right)^{\,\gamma} \cdot \nhi^{\,-\beta} \, ,
	\end{equation}
	where $\mathcal{N}_{\,0}$ is a normalization constant. This function defines the 1-dimensional distribution of the \hi{} present in the IGM probed by a random \los{}. The number $N_{\,abs}$ of systems for each \los{} is drawn from a Poisson distribution with parameter 
	\begin{equation} \label{eq:par}
 		\langle N_{\,abs} \rangle = \int_{\,I_{\,z}} \int_{\,I_{\,\nhi}} f\,(\nhi,z) \diff \nhi \diff z \, ,
	\end{equation}
	where the integral is carried out over appropriate redshift- and column density intervals $I_{\,z}$ and $I_{\,\nhi}$, respectively.\\
	We use different sets of input distributions that include the evolution of both low- and high density absorbers, and that give rise to the following models:

%__________________________________________________ table: Types of Abs.Syst. for the MMC model
   \begin{table}
      \caption[]{Types of absorbers and their corresponding parameters adopted from \citet[][their equation 10]{mad95a}. Note, however, that they quote $\nhi = 2.0 \cdot 10^{\,12}$ as lowest column density, while we use $\nhi = 1.0 \cdot 10^{\,12}$, in order for the adopted normalisation to be consistent.}
         \label{tab:typ1}
     $$ 
         \begin{array}{lccc}
            \hline
            \noalign{\smallskip}
            \nhi \ {[\mathrm{cm}^{\,-2}]}  &  \mathcal{N}_{\,0}  &  \gamma  & 
			\beta\\
            \noalign{\smallskip}
            \hline
            \noalign{\smallskip}
            10^{\,12} \ - \ 1.59 \cdot 10^{\,17}  &  2.40 \cdot 10^7  &  2.46  &  1.50  \\
            1.59 \cdot 10^{\,17} \ - \ 10^{\,20}  &  1.90 \cdot 10^8  &  0.68  & 
			1.50 \\
			\noalign{\smallskip}
            \hline
         \end{array}
     $$ 
%\begin{list}{}{}
%\item[$^{\,\mathrm{a}}$] This is footnote a
%\end{list}
   \end{table}
%__________________________________________________ end of table

%__________________________________________________ table: Types of Abs.Syst. for the BMC model
   \begin{table}
      \caption[]{Types of absorbers and their corresponding parameters, adopted from \citetalias[][their equation 10]{ber99a}.}
         \label{tab:typ2}
     $$ 
         \begin{array}{lccc}
            \hline
            \noalign{\smallskip}
            \nhi \ {[\mathrm{cm}^{\,-2}]}  &  \mathcal{N}_0  &  \gamma  & 
			\beta\\
            \noalign{\smallskip}
            \hline
            \noalign{\smallskip}
            10^{\,12} \ - \ 10^{\,14}  &  3.14 \cdot 10^7  &  1.29  &  1.46  \\
            10^{\,14} \ - \ 1.59 \cdot 10^{\,17}  &  1.70 \cdot 10^6  &  3.10  & 
			1.46 \\
            1.59 \cdot 10^{\,17} \ - \ 10^{\,20}  &  1.90 \cdot 10^8  &  0.68  & 
			1.50 \\
	\noalign{\smallskip}
            \hline
         \end{array}
     $$ 
%\begin{list}{}{}
%\item[$^{\,\mathrm{a}}$] This is footnote a
%\end{list}
   \end{table}
%__________________________________________________ end of table
%
%--------------------------------------------------------------------------------------------------------------
\paragraph*{MMC}

	This model relies on the input distributions from \citet[][their equation 10]{mad95a} listed in Table \ref{tab:typ1}. Here, the Doppler parameter is kept constant at a value \mbox{$b = 35.0 \kms$}, which roughly matches the mean $b$-value derived by \citet{rau92a}. This model corresponds to one of the several models presented by \citetalias{ber99a}, and is referred to as the \emph{MC-NH} method in their work.

%--------------------------------------------------------------------------------------------------------------
\paragraph*{BMC}

	This model matches the best method of \citetalias{ber99a} named \emph{MC-Kim}. The corresponding parameters for the line-density evolution and column density distribution functions are summarized in Table \ref{tab:typ2}.
	In this model, in contrast to the MMC model, the Doppler parameter for each absorber is drawn from a truncated, redshift-dependent Gaussian distribution of the form
	\begin{displaymath} \label{eq:par4}
 		P(b) \equiv \Theta\,(b - b_{\,tr}) \cdot \frac{1}{\sqrt{2 \pi \sigma^2}}
		\exp\left(-\frac{1}{2 \sigma^2} (b - \mu)^2\right) \, ,
	\end{displaymath}
	where $\Theta(x)$ is the Heaviside function:
	\begin{displaymath} \label{eq:par5}
 		\Theta(x) = \left\{ \begin{array}{ll}
					0, & x < 0\\
					1, & x \geq 0
					\end{array} \right.
	\end{displaymath}
	and the mean, standard deviation, and truncation value as a function of redshift are given by $\mu \, (z)= -3.85\,z + 38.9$, $\sigma \, (z)= -3.85\,z + 20.9$ and $b_{\,tr}\,(z) = -6.73\,z + 39.5$, respectively. \citetalias{ber99a} originally used this model to analyse the impact of the intergalactic attenuation on high-redshift galaxy colors in the range $1.75 < z < 5.0$, but we use it in the extended range $0.2 < z < 5.41$. Since our highest redshift limit is not too far away from \citetalias{ber99a}'s, we may safely assume that the model is valid in our extended redshift range.

%--------------------------------------------------------------------------------------------------------------
\paragraph*{MMC without Lyman limit systems}

	In order to asses the impact of the Lyman limit systems on the intergalactic absorption, we introduce this model, which consists of the same input distributions as the MMC model, excluding the systems with column densities $\nhi > 1.59 \cdot 10^{\,17} \cmsq$.\\

	In all the models describe above, the attenuation factor for each absorber is given by $\exp [-\tau \,(\lambda)]$, where, for the general case, the absorption coefficient $\tau \, (\lambda)$ can be written as
	\begin{equation} \label{eq:tau1}
 		\tau \, (\lambda) = \tau_{\,LL} \, (\lambda) + \sum_{\,i=2}^{\,N_{\,trans}}
		{\tau_i \, (\lambda)} \, .
	\end{equation}
	The first term on the right--hand side is the opacity due to the ionisation of neutral hydrogen by photons with wavelengths $\lambda \leq \lambda_{\,LL} \equiv 91.18$ nm. It is given by
	\begin{equation} \label{eq:tau2}
 		\tau_{\,LL} \, (\lambda) = \nhi \cdot \Theta\,(\lambda_{\,LL} - \lambda) \cdot g\,(\lambda) \cdot \sigma_{\,\infty} \cdot \left(
		\frac{\lambda}{\lambda_{\,LL}} \right)^3 \, ,
	\end{equation}
	where $\sigma_{\,\infty} \equiv 6.3 \cdot 10^{\,-18} \ \mathrm{cm}^2$ is the \hi{} photoionisation cross-section, and $g$ is the Gaunt--factor for bound--free transitions\footnote{An extensive tabulation of values for the Gaunt--factor can be found in \citet{kar61a}.}. The second term is the sum of the opacities due to resonant scattering at each transition of the Lyman series\footnote{We adopt the convention that the \lya{} transition (from the ground state to the next higher energy level) be identified with i = 2, the \lyb{} transition with i = 3, etc. The photoionisation cross-section is thus consistently denoted by $\sigma_{\,\infty}$.}. In general, the absorption coefficient for the transition $1 \to i$ is
	\begin{equation} \label{eq:tau3}
 		\tau_{\,i}(\lambda) = \nhi \cdot \sigma_{\,i} \cdot \phi \, (a_{\,i},x) \, .
	\end{equation}
	The cross-section $\sigma_{\,i}$ is a function of the Doppler parameter $b$, the oscillator strength of the transition $f_i$, and the resonant wavelength $\lambda_i$, and is given by
	\begin{equation} \label{eq:sig}
		\sigma_i = \frac{\sqrt{\pi}\,e^2}{m_{\,e} c^2} \frac{\lambda_i^2}{\Delta \lambda_D} \, f_{\,i} \, ,
	\end{equation}
	where $\Delta \lambda_D = \lambda_i \, b/c$ is the Doppler broadening, and the variable \mbox{$x \equiv (\lambda-\lambda_i)/\Delta \lambda_D$} is the distance to the line center in Doppler units. We assume the profile function $\phi$ of the absorption line to be given by the \VHf{}
	\begin{equation} \label{eq:vhf}
		H\,(a_{\,i},x) \equiv \frac{a_{\,i}}{\pi} \int_{-\infty}^{+\infty} \, \frac{\mathrm{e}^{-y^2}}{(x-y)^2 + a_{\,i}^{2}} \ dy \, .
	\end{equation}
	Here, \mbox{$a_{\,i} \equiv \lambda_i^2 \, \Gamma_i / (4 \pi \Delta \lambda_D)$} is the relative strength of the natural broadening to Doppler broadening for the $i$th transition, and \mbox{$y \equiv \vel/b$} is the kinetic velocity in units of the Doppler parameter. In this work, we neglect the opacity due to the photoionisation term and consider only the first resonant transition, \ie{} the \lya{} transition, since this is the only one of interest in the wavelength range studied here. Furthermore, we use the approximation to $H$ for values of $a$ and column densities characteristic for intergalactic \hi{} of \citet{tep06a}.

%--------------------------------------------------------------------------------------------------------------
\subsection{The transmission factor $\Phi$}
	
	The cumulative absorption along a random \los{} of the flux $f_{\,em}$ of a source at redshift $z_{em}$ is calculated according to expression
	\begin{equation} \label{eq:fobs}
 		f_{\,obs} \, (\lambda_{\,obs}) = \frac{f_{\,em} \, (\lambda_{\,em})}{1+\zem} \cdot
		\Phi \, (\lambda_{\,obs}) \, ,
	\end{equation}
	where $\lambda_{\,obs}$ and $\lambda_{\,em}$ are the observed and the emitted wavelengths, respectively. These are related by \mbox{$\lambda_{\,obs} = \lambda_{\,em} \cdot (1 + \zem)$}. The quantity $\Phi$ is the transmission factor and is given by
	\begin{equation} \label{eq:msk}
		\Phi \, (\lambda) \equiv \prod_{\,i=1}^{\,N_{\,abs}}\exp [ - \tau_{\,2} \, (\lambda/(1+z_{i})] = \exp\left[ - \sum _{\,i=1}^{\,N_{\,abs}} \tau_{\,2} \, (\lambda/(1+z_{i}) \right]\, ,
	\end{equation}
	where $\tau$ is given by equation (\ref{eq:tau3}), and $z_{\,i}$ is the redshift at the epoch of absorption\footnote{The reader shall bear in mind that $\nhi$, and $\sigma_{\,i}$ and $a_{\,i}$--through the dependence on the Doppler parameter--in equation (\ref{eq:tau3}) are different in general for each absorber. However, we do not write this explicitly by \eg{} introducing a new subscript in order to avoid a cumbersome notation.}.

	Introducing the relation $f_{\,c} \, (\lambda) = f_{\,em} \, [\lambda/(1+z)]/(1+z)$, it follows from equations (\ref{eq:cfd}), (\ref{eq:fobs}), and (\ref{eq:msk}) that
	\begin{equation} \label{eq:cfd2}
		1 - \da \, (z) =  \frac{1}{\Delta \lambda} \int_{\,\lambda_{\,1} \cdot (1+z)}^{\,\lambda_{\,2} \cdot (1+z)} \Phi \, (\lambda) \, \mathrm{d} \lambda \, .
	\end{equation}
	The right--hand side of this expression is just wavelength-averaged value of $\Phi$ at redshift $z$, and we will denote it by $\overline{\Phi}_{\,z}$. Note, however, that this quantity still depends on redshift, as indicated by the subscript. Since $\da \, (z)$ and $\overline{\Phi}_{\,z}$ differ only by a constant factor, they may be considered as equivalent with respect to their statistics, which will be discussed in the next section.

%---------------------------------------------------------------------------------------------------------------
\subsection{Distribution of $\da$} \label{sec:dda}

	Judging from the dependence of $\Phi$ on $\tau$ and $N_{\,abs}$ (see equation \ref{eq:msk}), it is expected that the transmission factor and consequently $\da$ are rather complicated random variables. Nevertheless, as is well known from statistics, a random variable $\bf x$ that can be expressed as the product of a large number of small, statistically independent factors is distributed lognormally, \ie{} according to the distribution
	\begin{equation} \label{eq:log}
		f(x;\mu,\sigma) = \frac{1}{\sqrt{2 \, \pi \, \sigma^{\,2}} \, x} \, \mathrm{e}^{\,-\frac{1}{2} \left(\frac{\ln x - \mu}{\sigma} \right)^{\,2}} \, ,
	\end{equation}
	where $\mu$ and $\sigma$ are the mean and the standard deviation of $\ln \bf x$. This expression is equivalent to the statement that a random variable $\bf x$ is distributed lognormally, if and only if its logarithm is distributed normally.

	The expected value $\mu'$ and the standard deviation $\sigma'$ of a lognormal distributed quantity can be expressed in terms of the parameters $\mu$ and $\sigma$ as
	\begin{equation} \label{eq:mul}
		\mu' = e^{\,\mu +\frac{1}{2} \sigma^{\,2}} \, ,
	\end{equation}
	and
	\begin{equation} \label{eq:sdl}
		\sigma' = \left(e^{\,\sigma^{\,2}} - 1\right)^{1/2} \, \mu' \, .
	\end{equation}
	From the form of equation (\ref{eq:msk}) we may suspect that the transmission factor is a lognormally distributed variable, since it can be expressed as the product of a large number of statistically independent factors that take on values in the range $[0, \, 1]$. The implications of this statement are profound: If $\Phi$ is distributed lognormally, so does  $\overline{\Phi}_{\,z}$ and consequently $\da \, (z)$. Furthermore, due to the property of the lognormal distribution stated above, the effective optical depth of the \lya{} absorption, usually defined as $\tau_{\,\mathrm{eff}} \equiv - \ln \, (1 - \da)$ \citep*[see \eg{}][]{kim01a} should obey a Gaussian distribution. This result follows independently from the fact that the total optical depth can be expressed as the sum of the independent contribution of each system, as indicated in equation (\ref{eq:msk}). Thus, for a sufficiently large number of absorbers $N_{abs}$, and if the optical depth for each absorber has the same mean value $\langle \tau \rangle$ and dispersion $\sigma\,(\tau)$ at each wavelength (\ie{} redshift), $\tau_{\,\mathrm{eff}}$ should obey a Gaussian distribution at each redshift, centered at $N_{abs} \,  \langle \tau \rangle$ and with a dispersion $\sqrt{N_{abs}} \,\sigma\,(\tau)$. These statements are completely independent of the form of evolution of the intergalactic neutral hydrogen, as long as the transmission factor can be expressed in the form of equation (\ref{eq:msk}).

	In order to test whether the values of $\da$ obey a lognormal distribution, we compute $\da$ using the MMC and BMC models for an ensemble of $N_{\,\mathrm{LOS}} = 4 \cdot 10^{\,3}$ \loss{} at a given redshift, and from these values we compute the mean and standard deviation of $\ln \da$ according to the equations
	\begin{equation} \label{eq:mea}
		\mu \, (\ln \da) \equiv \frac{1}{N_{\,\mathrm{LOS}}} \sum_{\,i=1}^{\,N_	{\mathrm	{LOS}}} \ln \da^{\,i} \, ,
	\end{equation}
	and
	\begin{equation} \label{eq:std}
		\sigma^{\,2}(\ln \da) \equiv \frac{1}{N_{\,\mathrm{LOS}}-1} \sum_{\,i=1}^{\,N_{\mathrm{LOS}}} \left(\ln \da^{\,i} - \langle \ln \da \rangle \right)^{\,2}\, ,
	\end{equation}
	With these parameters, we generate for each redshift a set of normally distributed random numbers using a standard random-number generator, and compare them statistically to the $\ln \da$ values from our simulations at each given redshift, in order to determine whether the latter are normally distributed. Recall that $\ln \da$ is distributed normally if and only if $\da$ obeys a lognormal probability distribution. As a first approach, we choose to compare both data sets via a $\chi^{\,2}$-test similar method, and to this end we compute for each set of values the 50 per cent quantile (median), and the $\pm 34.13, \pm 43.32, \pm 47.72, \pm 49.38$, and $\pm 49.87$ per cent quantiles around the median. We define the following measure
	\begin{figure}
		\begin{center}
		\resizebox{\hsize}{!}{\includegraphics{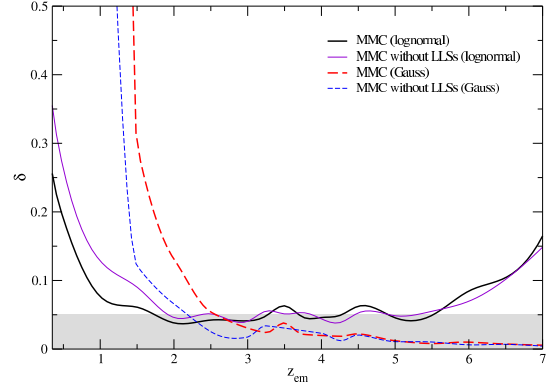}}
		\caption{Values of $\delta$ for the MMC model assuming a lognormal (heavy solid line) and Gaussian (heavy dashed line) parent distribution. The shaded area indicates the range $\delta \leq 0.05$, that corresponds to a probability of $1.9\cdot10^{-5}$ that the data are not drawn from the same parent distribution. Note how the assumption of a Gaussian parent distribution improves with redshift, but that a lognormal distribution is a better assumption for low redshifts. The light solid and light dashed lines correspond to the MMC model without Lyman limit systems (see Section \ref{sec:lls}). Note that the approximation to a lognormal distribution, is better when these systems are included (see text for discussion). The fluctuations seen with respect to a perfect smooth curve in each case are due to the random nature of the process at each redshift, and are not significant.}
		
		\label{fig:del}
		\end{center}
	\end{figure}
	\begin{equation} \label{eq:del}
		\delta^{\,2} \equiv \sum_{\,i=1}^{\,N_{\,q}} \left(\frac{Q_{\,i} - q_{\,i}}{Q_{\,i}}					\right)^{\,2} \, ,
	\end{equation}
	where the $Q$'s are the quantiles determined from the set of random numbers distributed normally, the $q$'s are the corresponding quantiles of the $\ln \da$ values at a given redshift that result from our simulations, and $N_{\,q} = 11$ is the number of quantiles. By definition, $\delta$ explicitly gives the absolute deviation of one data set as characterised by these 11 quantiles with respect to the other data set, and thus quantifies the departure of the assumed distribution. Indeed, the smaller the value of $\delta$, the larger the probability that both data sets belong to the same parent distribution. The values of $\delta$ computed in this way turn out to lie in the range $(5 \cdot 10^{\,-2}, \, 5 \cdot 10^{\,-1})$, which may seem vanishingly small. However, since the distribution of the random variable $\delta$ itself is unknown, we determine the significance of the results by generating two sets of normally distributed random numbers with mean and standard deviations distributed uniformly, and comparing them to the values found for our simulated data. We do this for $1.2\cdot10^{\,6}$ pairs of sets, and determine from these what is the fraction of realisations with a value of $\delta$ smaller than a given value. According to these estimates, there is a probability of $\{1.4\cdot10^{-2},3.1\cdot10^{\,-3}, \, 1.9\cdot10^{\,-4}, \,  2.3\cdot10^{\,-5}\}$ that a value $\delta \leq \{0.50, \, 0.25, \, 0.10, \, 0.05\}$ will be produced by chance, respectively.  In Figure \ref{fig:del} we show the values for $\delta$ as a function of redshift for the full MMC model, assuming a lognormal parent distribution for $\da$ (heavy solid line). The result are qualitatively the same for the BMC model and are not shown for clarity. Note that for the whole redshift range shown, $\delta \lesssim  0.5$, and in particular $\delta \lesssim 0.1$ for $\zem > 1.5$.
	\begin{figure*}
		\begin{center}
			\resizebox{\hsize}{!}{\includegraphics{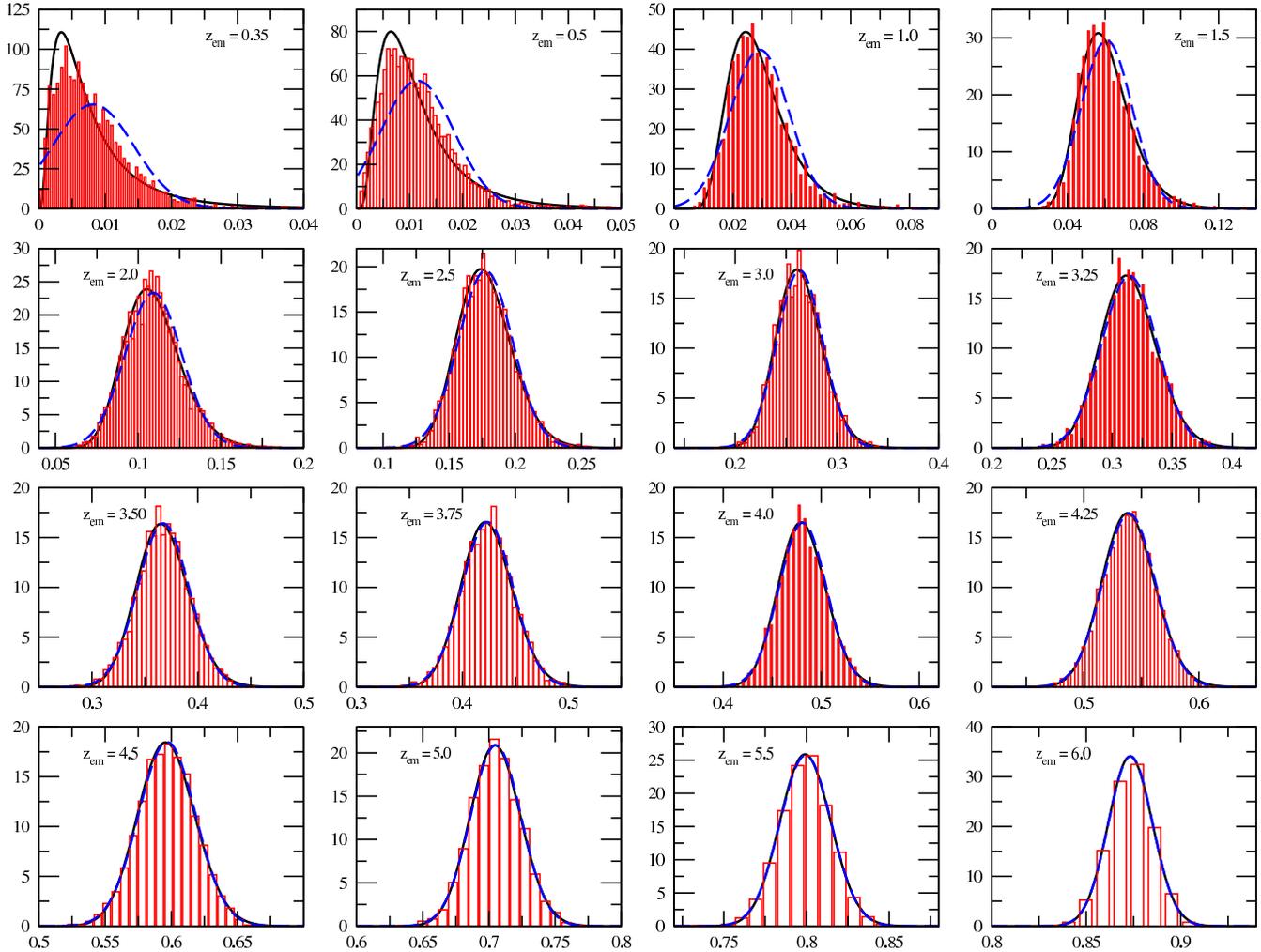}}
			\caption[Flux distribution]{Distribution of the $\da$ values at a given redshift computed from simulations based on the MMC model (histogram). The distribution is normalised to unit area. The bin size used to compute the histogram at each redshift has been arbitrarily chosen to be given by $max\{\,\da\,(z)\} / 100$, and is hence different at each redshift. The binning of the data, \ie{} the $\da$ values at each redshift obtained from the simulations, is intended only for display purposes. Shown are also the analytic lognormal distribution (solid line) computed according to equations (\ref{eq:log}), (\ref{eq:mea}), and (\ref{eq:std}), and the analytic Gaussian distribution (dashed line) computed using the same equations as before but replacing $\ln \da$ by $\da$. The $y$-axis indicates the probability of the corresponding $\da$ value on the $x$-axis. Note that the binned data have not been used in any form to estimate the parameters of the corresponding probability distribution. Hence, the curves shown are not fits to the binned data, but are computed using only the mean and standard deviation of the raw data. Note the excellent agreement between the data and the analytic lognormal distribution, especially with respect to the skewness and the cut-off at $\da = 0$. The distribution of the data is not well described by a Gaussian distribution at the lowest redshifts shown. However, the approach of the data distribution to a Gaussian distribution increases with increasing redshift, as expected (see text for details).}
		\label{fig:fdd}
		\end{center}
	\end{figure*}

	It follows from the Central Limit Theorem that the larger the number of absorbers $N_{\,abs}$, the closer the approach of the distribution of $\Phi$ to a lognormal distribution, since the distribution of $\ln \Phi$ approaches a Gaussian. Given that $N_{\,abs}$ increases with redshift, it should be expected that the accuracy with which the distribution of $\Phi$ approaches a lognormal distribution also increases with redshift. Furthermore, if the integral in equation (\ref{eq:cfd2}) is approximated by a sum of the form
	\begin{displaymath}
		\overline{\Phi}_{\,z} \approx \frac{1}{N} \sum_{i=1}^{N} \Phi_{\,i} \, \Delta \lambda_{\,i} \, ,
	\end{displaymath}
	where $N$ is the number of pixels, it is apparent that for a sufficiently large $N$, and assuming that $\Phi$ has the same mean value $\langle \Phi \rangle$ and dispersion $\sigma(\Phi)$ at each pixel, the distribution of $\overline{\Phi}_{\,z}$ should approach a Gaussian distribution with mean $\langle \Phi \rangle$ and dispersion $\sigma(\Phi) / N$. Since the width of a given restframe wavelength range, \ie{} the number of pixels also increases with redshift as $(1+z)$, it is expected that the approximation of the distribution of $\overline{\Phi}_{\,z}$ and hence of $\da$ to a Gaussian becomes better with increasing redshift. We thus have the superposition of two effects: On the one hand, the distribution of $\Phi$ at a fixed wavelength approaches a lognormal distribution at any given redshift, with the accuracy increasing with redshift. On the other hand, the distribution of $\overline{\Phi}_{\,z}$ and $\da\,(z)$ approaches a Gaussian distribution with increasing redshift. The net result should be that $\overline{\Phi}_{\,z}$ and $\da$ are distributed lognormally at low redshifts, and that their distribution approaches a Gaussian for higher redshifts. Since $\da$ asymptotically converges to unity for very high redshifts, its distribution at these redshifts is expected to be highly peaked around the mean value. This is naturally given by the fact that the dispersion of $\overline{\Phi}_{\,z}$ scales like $\sigma(\Phi) / N$ around $\langle \Phi \rangle$, and that $N$ increases with redshift, as stated above.

	In order to test this expectation, we compute the analytical lognormal (Gaussian) probability distribution of $\da$ as given by equation (\ref{eq:log}) with the mean and standard deviation of $\ln \da$ ($\da$) computed using equations (\ref{eq:mea}) and (\ref{eq:std}) from the ensemble of values that result from our simulations at each redshift. The comparison of the analytic distribution thus obtained to the distribution of simulated $\da$ values is shown for the full MMC model in Figure \ref{fig:fdd}. The results for the BMC model are qualitatively the same and are hence not shown. Note the excellent agreement at all redshifts between the lognormal probability distribution and the distribution of the $\da$ values resulting from our simulations. We want to emphasize that the curves shown are not fits to the binned data, but are computed using only the mean and standard deviation of the unbinned data. At lower redshift, the agreement at the lower cut-off, \ie{} at $\da = 0$ is worth mentioning. It is remarkable that this cut-off, which is physically given by the fact that absorption as measured by $\da$ cannot take on values smaller than zero, arises in a natural way due solely to the fact that $\da$ is distributed lognormally. Note, in contrast, that a Gaussian distribution does not satisfactorily describes the distribution of the data at these low redshifts; in particular, the Gaussian distribution does not display the sharp cut-off at $\da = 0$. Nevertheless, the description of the data by a Gaussian distribution becomes better with increasing redshift, as expected, and also the corresponding Gaussian distribution becomes narrower at every increasing redshift (taking into account the change in the scale of the $x$-axis from the first to the last panel). Moreover, note that the analytic lognormal and Gaussian distribution become visually indistinguishable from each other at $\zem > 3.0$. In order to assess quantitatively the differences between these distributions with respect to the distribution of the data, we compute again the value of $\delta$ at each redshift, assuming now that the $\da$ values obtained from the MMC model are drawn from a Gaussian parent distribution. We compare these values to the corresponding values computed before for an assumed lognormal parent distribution. This comparison is shown in Figure \ref{fig:del}. As can be seen, the values of $\delta$ for both distributions are low at all redshifts, and the difference between them at $\zem \gtrsim 2$ is negligible small. This explains why the lognormal and Gaussian distribution functions shown in Figure \ref{fig:fdd} are practically indistinguishable from each other. However, note that the values of $\delta$ for a assumed Gaussian parent distribution become vanishingly small with increasing redshift, eventually becoming smaller than the corresponding values for the lognormal distribution. Besides, it can be see also that the $\delta$ values for a assumed lognormal distribution rise again towards high redshifts, implying that the distribution of the data at these redshifts is no longer well described by a lognormal distribution.This confirms our statement made above, that $\da$ is expected to be distributed lognormally at low redshifts and normally at higher redshifts. The redshift at which the transition from a lognormal to a Gaussian distribution takes place may depend on the particular set of input distributions used. For redshifts where the distribution of $\da$ is well approximated by a lognormal distribution, the optical depth should be distributed normally. This result is however not in agreement with the results from \eg{} \citet[][his Figure 1]{mad95a}, \citet*[][their Figure B1]{mei04a}, or \citet[][their Appendix C]{ber03a}, where the distribution of the mean flux, \ie{} the averaged transmission factor $\overline{\Phi}_{\,z}$ is claimed to be highly inconsistent with a Gaussian distribution. It is usually assumed that the presence of the high-density Lyman limit systems is responsible for this behaviour, and we next make use of our simulations to address this question.

%---------------------------------------------------------------------------------------------------------------
\subsubsection{The Effect of the Lyman limit systems (LLSs)} \label{sec:lls}

	In order to quantify the effect that the LLSs have on the distribution of $\da$, we compare, following the analysis of Section \ref{sec:dda}, the results of the simulations for the MMC model with and without the optically thick LLSs.
	We find that $\da$ is lognormally distributed as well with a high confidence for the case where the LLSs are excluded, as can in Figure \ref{fig:del}, where the values for $\delta$ computed at each given redshift are shown. Again, the results for the BMC model are qualitatively the same and are omitted here. Note that the values of $\delta$ for the MMC model without LLSs are larger at redshifts $\zem \lesssim 3$ than those for the full MMC model. This is due to the fact that the absence of the optically thick LLSs enhances the probability of the attenuation factor $\exp \, (-\tau)$ to be closer to unity at a given wavelength along a random \los{}. This means effectively that the number of factors in equation (\ref{eq:msk}) are reduced. Hence, when the LLSs are absent, the approach to a lognormal distribution should be worse with respect to the case where the LLSs are included. Furthermore, this effect should be enhanced towards lower redshifts, for which the number of factors, \ie{} of absorbers decreases as $\propto (1+z)^{\,\gamma}$. Note that the trend is the opposite in the case of a assumed Gaussian parent distribution.
	\begin{figure}
		\begin{center}
		\resizebox{\hsize}{!}{\includegraphics{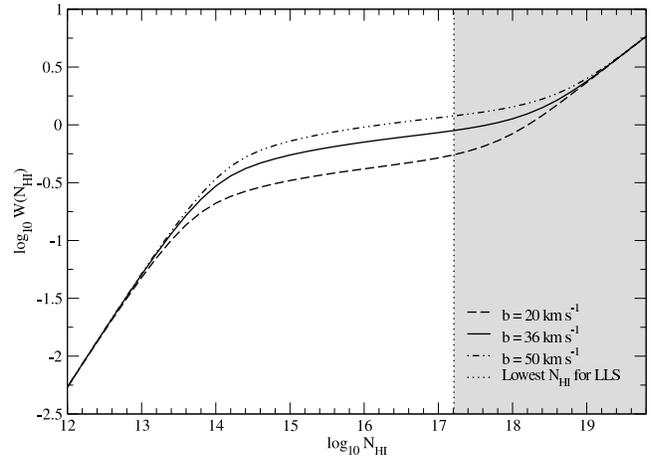}}
		\caption[]{Curve of growth of the \lya{} absorption line, for three typical values of the Doppler parameter. The shaded region corresponds to the column densities characteristic to Lyman limit systems.}
		\label{fig:cog}
		\end{center}
	\end{figure}

	We hence find when computing the evolution of $\da$ with the MMC model with and without LLSs that the predictions for the evolution of $\da$ with redshift for the full MMC model are practically indistinguishable from the results of the MMC model without LLSs, which implies that the effect of these systems on the total absorption is small. Furthermore, it turns out that the distribution of $\da$ is  lognormal irrespective of the presence of LLSs. All these results point to the fact that the LLS have a negligible impact on the evolution of $\da$, as long as the input distributions used here correctly describe the number density evolution of the absorbers. This would confirm similar results already found by \citet*{des07a}. Also, as pointed out by  \citet{mcd05a}, the picture would not be fundamentally different if higher column density systems such as damped \lya{} systems were present. We will come back to this point in Section \ref{sec:res}.\\

	We think that, as long as the distribution functions realistically describe the evolution of the \lya{} absorbers, the LLSs \emph{cannot} truly have a great impact neither on the absorption as measured by $\da$, nor on its statistics because of the following reasons: 1) They are scarce, and even more compared to the thinner \lya{} forest systems (\eg{} of the order of 5 LLSs on average along a random \los{} compared to approximately $10^{\,3}$ \lya{} forest absorbers out to $z = 3.0$, according to the MMC model); 2) Their contribution to the absorption is due solely to the absorption at the resonant wavelength of the \lya{} line, and the equivalent width of the \lya{} absorption line arising in systems with column densities $\nhi = 10^{\,17.21 - 18} \cmsq$ is not too different from that of the systems with column densities $\nhi = 10^{\,14 - 15} \cmsq$ (cf. the curve-of-growth in Figure \ref{fig:cog}), which are by far more numerous. In order to estimate the relative contribution of each population at a given redshift, we weight the equivalent width $W(N,b)$ of the \lya{} absorption line as a function of the column density $N$ and Doppler parameter $b$ with the column density distribution, and compute the ratio
\begin{equation}
	\varrho \, (\zem;b) \equiv \frac{\langle W \rangle_{\,\mathrm{Ly\alpha}}}{\langle W \rangle_{\,\mathrm{LLS}}} \, ,
\end{equation}
	where
	\begin{equation}
		\langle W \rangle_{\,i} \equiv \int_{\,0}^{\zem} \int_{\,N_{\,min}}^{\,N_{\,max}} \, (1+z) \cdot W_{\,0}\,(N,b) \, f _{\,i}\,(N,z) \diff N \diff z \, ,
	\end{equation}
	Here, $f_{\,i}$ is the distribution function of population \mbox{$i \in \{\mathrm{Ly\alpha}, \, \mathrm{LLS}\}$}, $W_{\,0} \, (N,b)$ is the rest equivalent width of a \lya{} absorption line for a column density $N$ and Doppler parameter $b$, and $N_{\,min}$ and $N_{\,max}$ are the column density limits that define each population, respectively. For a reasonable value for the Doppler parameter of $36 \kms$, and the input distributions of the MMC model, we find that $\varrho >> 1$ at all redshifts. The result is qualitatively the same for the BMC model. Hence, It follows from these models that the \lya{} forest systems dominate the absorption over the optically thicker Lyman limit systems at all epochs. This would explain at least in part why the difference between the predictions for $\da$ from the MMC model with and without LLSs, is small. Also, it is consistent with the theoretical expectation that the distribution of $\da$ should not be far from lognormal or Gaussian, with or without LLSs, since this only depends on the fact that the absorption factor be expressed in the form of equation (\ref{eq:msk}), and this is truly independent of the form of the input distributions, as stated previously.
%---------------------------------------------------------------------------------------------------------------
\subsection{Scatter in $\da$} \label{sec:sca}

	We expect the intrinsic scatter in the absorption due to cosmic variance to be strongest at \loss{} of middle length. At low redshifts, both the number of thin \lya{} forest clouds and thick Lyman limit systems is small, and the addition of a few more does not change dramatically the amount of absorption. However, the mean number of \lya{} forest clouds and LLS increases with $z$, and thus the probability of encountering more or less systems than average increases as well. Correspondingly, the absorption increases and so does its scatter. At even higher redshifts, the number of \lya{} forest systems increases so dramatically and the absorption is so severe that the addition of more systems does not make any difference neither to the absorption nor to the scatter. Thus, we should expect the stochastic effect, \ie{} the scatter in absorption, to peak at some intermediate redshift $z_{\,int}$.
	\begin{figure}
		\begin{center}
			\resizebox{\hsize}{!}{\includegraphics{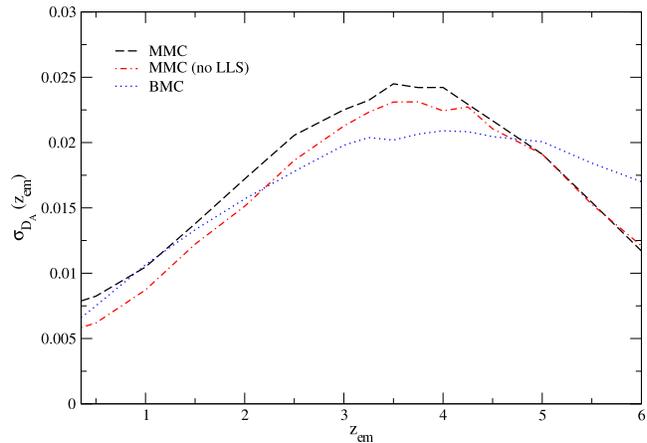}}
			\caption[]{Evolution of the intrinsic scatter of $\da$ due to the stochastic nature of the absorption in the intergalactic medium for the two competing models MMC (dashed line) and BMC (dotted line). The  dot-dashed line corresponds to the MMC without Lyman limit systems (see Section \ref{sec:lls} for details). Note that the behaviour of $\sigma(\da)$ is qualitatively the same for all three models. As expected, the amount of scatter for the MMC model is less in the case where the Lyman limit systems are absent.}
		\label{fig:sca}
		\end{center}
	\end{figure}

	We compute the intrinsic scatter at the $\sigma$ level for the values of $\da$ at each given redshift obtained from our simulations using equations (\ref{eq:mul}), (\ref{eq:sdl}), (\ref{eq:mea}), and (\ref{eq:std}). The result is shown in Figure \ref{fig:sca}, where we show the evolution of the scatter in $\da$ with redshift for the MMC model with and without LLSs. We also include for comparison the result from the BMC model. It can be seen that, irrespective of the model, the scatter peaks at a intermediate redshift between $\zem \approx 3.5$ and $\zem \approx 4.0$. Note that the peak is significant, since it represents an increase in the scatter of 2.5 times with respect to its value at $\zem \approx 1$. It is interesting that this result had also been found by \citet*[][their Figure 2]{zuo93a}, who using a semi-analytic approach and different input distributions, reported that $\sigma \, (\da)$ is largest at redshifts near 3.7. Thus, the qualitative behaviour of the intrinsic scatter of $\da$ shown in Figure \ref{fig:sca} may be an unavoidable feature of this observable, which could explain at least in part the large scatter in the measurements of $\da$ seen in the same redshift interval (cf. Figure \ref{fig:daz}). As will be shown in Section \ref{sec:res}, however, there is a disagreement at the $3\,\sigma$ level between the \emph{amplitude} of the scatter in the observations at these redshifts and that predicted by the models.

	When comparing models that only differ by the presence of the optically thick Lyman limit systems, we find that the scatter in $\da$ is larger at any given redshift when the LLS are present, as expected. None the less, the absolute value of the scatter does not differ significantly between the situation where these systems are present and where they are absent. Thus, the net effect of the LLSs is to enlarge the intrinsic scatter in the absorption, contributing only marginally to the mean value of the absorption itself. However, note that for redshifts $\zem \gtrsim 4.0$, the amount of scatter at the 1$\sigma$ level becomes indistinguishable between both scenarios, owing to the fact that the thinner \lya{} forest systems are overwhelmingly more numerous at these redshifts.

	The results stated above are consistent with the previous conclusion obtained from our simulations that the optically thick LLSs cannot have a great impact on the absorption as measured by $\da$ and on its scatter.
%--------------------------------------------------------------------------------------------------------------
\section{Our $\da$ Measurements} \label{sec:mda}
	
	We want to compare the predictions for the evolution of $\da$ that result from the models described above to observations. For this purpose, we use previous measurements of $\da$ reported in the literature, and we perform ourselves new measurements of this quantity, using QSO spectra from the SDSS Data Release 5 \citep{ade07a}, thus extending the redshift range of the measurements to $\zem = 5.41$.

	Our selection procedure of sources suitable for this purpose was as follows: Since the wavelength range available from SDSS (DR5) is $\lambda \in [380,920] \nm$, and we measure the continuum depression $\da$ in the restframe wavelength interval $\lambda \in [105,117] \nm$, the redshift of our sample is restricted to $z_{\,min} \geq 380/105 - 1 = 2.62$. We choose $z_{\,min} = 2.7$ as our lowest redshift in order to avoid the low S/N at the blue end of the spectrograph. A simple query for high-z quasi-stellar objects on the SDSS SkyServer Spectroscopic Query Form with this restriction alone returns around 2400 spectra. From this first selection, we rejected those objects for which the redshift was either not measured, the measurement had failed, or the measured photometric and spectroscopic redshifts were inconsistent with each other. We binned the quasars in redshift intervals of $\Delta z = 0.1$ and selected for each redshift bin the spectrum with the highest S/N, leaving us with 28 sources, from which we removed three further objects due to low data quality. The resulting sample is listed in Table \ref{tab:qso}.

\subsection{Continuum fit} \label{sub:con}
	
	The continuum of a quasar is often assumed to be of the form \mbox{$f_{\,\nu} = f^{\,0}_{\,\nu} \cdot \nu^{\,-\alpha_{\,\nu}}$} \citep[see \eg{}][]{ste87a,lao97a}, or equivalently, \mbox{$f_{\,\lambda} = f^{\,0}_{\,\lambda} \cdot \lambda^{\,-\alpha_{\,\lambda}}$}, where both indices are related by \mbox{$\alpha_{\,\lambda} = 2 - \alpha_{\,\nu}$}, and \mbox{$f^{\,0}_{\,\lambda} = f^{\,0}_{\,\nu} \cdot c^{\,1-\alpha_{\,\nu}}$}, and $c$ is the speed of light. Empirical evidence in favour of the assumption of an underlying continuum in form of a power-law has been given by \eg{} \citet*{obr88a}, who derived an average power-law index of $\alpha_{\, \nu} = 2.36$ for \mbox{$\lambda \in [80, \,121.6) \nm$} and $\alpha_{\, \nu} = 0.67$ for \mbox{$\lambda \in [121.6, \,190] \nm$}. This is consistent with the result already found by \citet{neu79a} that the spectral index $\alpha_{\,\lambda}$ varies over large wavelength ranges. \citet{zhe97a} found by constructing a composite QSO spectrum from 284 HST FOS spectra that a single power-law describes well the continuum for wavelengths between 105 and 220 nm with $\alpha_{\, \nu} = 0.99 \pm 0.05$, but that the continuum steepens (flattens in $\lambda$) significantly for $\lambda \leq 105 \nm$ ($\alpha_{\, \nu} = 1.96 \pm 0.15$). Similarly, \citet{tel02a} reported, using a sample nearly twice as large as the \citeauthor{zhe97a}'s sample, that the continuum in the extreme UV region between $\lambda = 50 \nm$ and $\lambda = 120 \nm$ and in the near UV between $\lambda =  120 \nm$ and $\lambda = 300 \nm$ is well described by a single power-law with $\alpha_{\, \nu} = 1.76 \pm 0.12$ and $\alpha_{\, \nu} = 0.69 \pm 0.06$, respectively. Furthermore, by constructing a composite QSO spectrum from a homogeneous sample of over 2200 SDSS QSOs, \citet{van01a} concluded that the continuum in the rest-frame wavelength range $121.6 < \lambda < 500 \nm$ can be very well modeled by a single power-law either in wavelength or frequency, with $\alpha_{\,\nu} = 0.44 \pm 0.1$. However, they did not estimate a spectral index for  wavelengths $\lambda < 121.6 \nm$. While it is the consensus that the optical/NIR continuum of quasar spectra is due to synchrotron emission \citep*{whi01a}, the physical origin of the quasar UV/optical continuum is not yet established. On the hand hand, the observed soft X-ray/UV/optical spectral shape of quasars is found to be consistent with free-free emission of optically thin gas at temperatures $10^{\,5} - 10^{\,6}$ K \citep{bar93a}, which is well described by a power-law. On the other hand, it has been argued that the emission from a geometrically thin, optically thick accretion disk also reproduces well the observed quasar SED in this spectral region \citep*{kaw01a}. Either model, however, is consistent with an underlying continuum in the form of a power-law. Alternatively, as already mentioned in Section \ref{sec:cfd}, other authors prefer to fit quasar continua locally using spline functions \citep*[see \eg{}][]{lu96a,kir03a,tyt04a,tyt04b} searching for regions apparently free of absorption blueward of the \lya{} emission line.

	Since the \lya{} forest region is severely absorbed due to intervening \hi{} systems, especially for high-z QSOs, and a local fit to the continuum in this region is difficult, with the uncertainty increasing with redshift \citep[see \eg{}][]{fau07a}, we choose to estimate the continuum of our selected sources in the \lya{} forest region by fitting a power-law to the QSO spectrum redward of the \lya{} emission line and extrapolating it for $\lambda \leq 121.567 \nm$. As long as the assumption of the underlying power-law holds, this approach has the advantage that the continuum estimate is completely independent of the spectral resolution and S/N in the \lya{} forest region. Nevertheless, this method may lead to an systematic overestimate of the continuum flux and thus to an corresponding overestimate of $\da$. It follows from the results quoted above that the quasar's intrinsic continuum in spectral regions below and above the \lya{} emission line may both be well described by power-laws, but with potentially different spectral indices. More specifically, the spectrum in the region below 121.6 nm (rest-frame) may be steeper in frequency, and hence flatter in wavelength\footnote{Recall that  $\alpha_{\,\lambda} = 2 - \alpha_{\,\nu}$, which implies $\diff \alpha_{\,\lambda} = - \diff \alpha_{\,\nu}$.}, with respect to the continuum in the region above 121.6 nm \citep*[see \eg][their Figure 4]{tel02a}. Thus, fitting a continuum in the form of a power-law at wavelengths $\lambda > 121.6 \nm$ and extrapolating it towards $\lambda < 121.6 \nm$ may well lead to an overestimate of the continuum level in that region, as first pointed out by \citet*{sel03a}, according to whom the $\da$ may be overestimated by at least 0.05 at $\zem = 2.72$. We will take this systematic bias into account when estimating the uncertainty in our $\da$ measurements (cf. Section \ref{sub:dam}).
%
%--------------------------------------------------------------------------------------------------------------
\begin{table*}
\caption{QSO sample selected from the SDSS DR5. The first four columns include the object designation$^\mathrm{\,a}$, the emission redshift as quoted in the SDSS DR5 catalog, the spectral index used for fitting the continuum, and its uncertainty, respectively. The last three columns list the measurement of $\da$ and its total uncertainty, and the uncertainty due to different error sources, respectively (see text for details.)}
\begin{center}
\begin{tabular}{lcccccc}
Object & $\zem$ & $\alpha_{\,\lambda}$ & $\sigma_{\,\mathrm{fit}} \, (\alpha_{\,\lambda})$ & $\da$ & $\sigma_{\,\mathrm{fit}} \, (\da)$ & $\sigma_{\,\mathrm{sys}} \, (\da)$\\[4pt]
\hline
SDSS J115538.60+053050.5 & 2.712 & -1.213 & 0.002 & $0.390^{+0.010}_{-0.040}$ & $^{+0.010}_{-0.010}$ & 0.039\\[6pt]
SDSS J112107.99+513005.4 & 2.843 & -1.256 & 0.003 & $0.281^{+0.015}_{-0.032}$ & $^{+0.015}_{-0.016}$ & 0.028\\[6pt]
SDSS J010619.24+004823.3 & 2.882 & -1.170 & 0.003 & $0.294^{+0.018}_{-0.035}$ & $^{+0.018}_{-0.018}$ & 0.029\\[6pt]
SDSS J075618.13+410408.5 & 2.956 & -1.320 & 0.007 & $0.329^{+0.038}_{-0.052}$ & $^{+0.038}_{-0.040}$ & 0.033\\[6pt]
SDSS J164219.89+445124.0 & 3.125 & -1.550 & 0.001 & $0.349^{+0.005}_{-0.035}$ & $^{+0.005}_{-0.005}$ & 0.035\\[6pt]
SDSS J004054.65-091526.8 & 3.185 & -1.278 & 0.000 & $0.459^{+0.000}_{-0.046}$ & $^{+0.000}_{-0.000}$ & 0.046\\[6pt]
SDSS J124306.55+530522.1 & 3.317 & -1.255 & 0.012 & $0.426^{+0.057}_{-0.077}$ & $^{+0.057}_{-0.064}$ & 0.043\\[6pt]
SDSS J083122.57+404623.4 & 3.365 & -1.230 & 0.003 & $0.261^{+0.017}_{-0.031}$ & $^{+0.017}_{-0.017}$ & 0.026\\[6pt]
SDSS J085343.32+370402.3 & 3.475 & -1.115 & 0.005 & $0.527^{+0.019}_{-0.056}$ & $^{+0.019}_{-0.020}$ & 0.053\\[6pt]
SDSS J093523.32+411518.7 & 3.566 & -1.407 & 0.007 & $0.396^{+0.036}_{-0.055}$ & $^{+0.036}_{-0.039}$ & 0.040\\[6pt]
SDSS J094349.65+095400.9 & 3.713 & -1.199 & 0.014 & $0.529^{+0.057}_{-0.083}$ & $^{+0.057}_{-0.064}$ & 0.053\\[6pt]
SDSS J023137.64-072854.5 & 3.750 & -1.269 & 0.008 & $0.462^{+0.037}_{-0.061}$ & $^{+0.037}_{-0.039}$ & 0.046\\[6pt]
SDSS J144717.97+040112.4 & 3.931 & -1.303 & 0.016 & $0.525^{+0.062}_{-0.089}$ & $^{+0.062}_{-0.072}$ & 0.052\\[6pt]
SDSS J162331.15+481842.1 & 3.990 & -1.113 & 0.006 & $0.494^{+0.022}_{-0.055}$ & $^{+0.022}_{-0.023}$ & 0.049\\[6pt]
SDSS J014049.18-083942.5 & 4.112 & -1.181 & 0.010 & $0.488^{+0.041}_{-0.066}$ & $^{+0.041}_{-0.044}$ & 0.049\\[6pt]
SDSS J234150.01+144905.9 & 4.155 & -1.094 & 0.010 & $0.558^{+0.035}_{-0.067}$ & $^{+0.035}_{-0.038}$ & 0.056\\[6pt]
SDSS J081240.68+320808.6 & 4.332 & -1.074 & 0.028 & $0.517^{+0.100}_{-0.137}$ & $^{+0.100}_{-0.127}$ & 0.052\\[6pt]
SDSS J103601.03+500831.8 & 4.449 & -1.149 & 0.012 & $0.612^{+0.039}_{-0.075}$ & $^{+0.039}_{-0.043}$ & 0.061\\[6pt]
SDSS J162626.50+275132.4 & 4.580 & -1.187 & 0.023 & $0.681^{+0.057}_{-0.097}$ & $^{+0.057}_{-0.069}$ & 0.068\\[6pt]
SDSS J005006.35+005319.2 & 4.663 & -1.204 & 0.018 & $0.693^{+0.043}_{-0.085}$ & $^{+0.043}_{-0.049}$ & 0.069\\[6pt]
SDSS J083914.14+485125.7 & 4.885 & -1.350 & 0.030 & $0.737^{+0.057}_{-0.104}$ & $^{+0.057}_{-0.073}$ & 0.074\\[6pt]
SDSS J163950.52+434003.7 & 4.976 & -1.321 & 0.024 & $0.694^{+0.056}_{-0.098}$ & $^{+0.056}_{-0.069}$ & 0.069\\[6pt]
SDSS J233446.40-090812.3 & 5.107 & -1.332 & 0.000 & $0.735^{+0.000}_{-0.073}$ & $^{+0.000}_{-0.000}$ & 0.073\\[6pt]
SDSS J101447.18+430030.1 & 5.275 & -1.220 & 0.029 & $0.758^{+0.058}_{-0.107}$ & $^{+0.058}_{-0.076}$ & 0.076\\[6pt]
SDSS J142123.98+463317.8 & 5.414 & -1.321 & 0.080 & $0.825^{+0.087}_{-0.192}$ & $^{+0.087}_{-0.173}$ & 0.082\\[6pt]
\noalign{\smallskip}
\end{tabular}
\end{center}
\label{tab:qso}
\begin{list}{}{}
\item[$^{\,\mathrm{a}}$] The designation of each object meets the IAU nomenclature, as required. For details on the official SDSS designation of an object, please consult www.sdss.org/dr5/coverage/IAU.html
\end{list}

\end{table*}
%
%--------------------------------------------------------------------------------------------------------------
	
\begin{figure}
		\begin{center}
			\resizebox{\hsize}{!}{\includegraphics{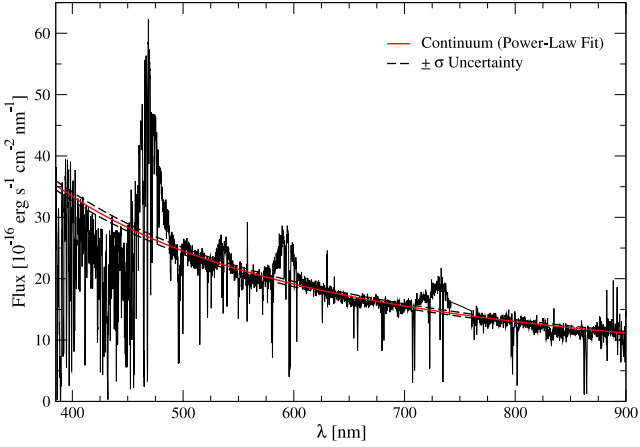}}
			\caption[]{Spectrum of the QSO SDSS J112107.99+513005.4 at \mbox{$\zem = 2.843$}. The heavy solid and heavy dashed line indicate the power-law fit to the continuum and its uncertainty, respectively. The corresponding spectral index is \mbox{$\alpha_{\,\lambda} = 1.256\pm 0.003$}.}
		\label{fig:qso}
		\end{center}
	\end{figure}

	The power-law that we fit to each spectrum is of the form
\begin{equation} \label{eq:pow}
f_{\,c} \,(\lambda;\alpha_{\,\lambda}) = f^{\,0}_{\,\lambda} \cdot \left(\lambda + \lambda_{\,0} \right)^{\,-\alpha_{\,\lambda}} \, .
\end{equation}
	where the flux amplitude $f^{\,0}_{\,\lambda}$, the wavelength off-set $\lambda_{\,0}$, and the spectral index $\alpha_{\,\lambda}$ are the parameters to be determined. We fit the continuum in the wavelength range \mbox{$[130\cdot(1+\zem), \, 900] \nm$}, in order to avoid the red emission wing of the \lya{} line and the red end of the spectrograph, respectively. Note that the available wavelength range decreases with redshift, and this may introduce an increasing uncertainty in the fitted continuum. The fit parameters and their uncertainties were obtained with help of the IDL task CURVEFIT. It turns out that the continuum fit is rather insensitive to the uncertainties in the flux amplitude and the wavelength off-set, and extremely sensitive to the uncertainty in the spectral index. Because of this and for simplicity, in our further analysis we neglect the error in the first two parameters and consider only the uncertainty in the spectral index to be of relevance\footnote{The full list of fit parameters for each source and their uncertainty are available in machine-readable form at http://astro.physik.uni-goettingen.de/\~{}tepper/da/fitparam.txt}. An example of a QSO spectrum and its corresponding fit are shown in Figure \ref{fig:qso}. The spectral index and its uncertainty for each source are listed in Table \ref{tab:qso} (columns three and four, respectively). The uncertainty in the spectral index quoted in column four and denoted by $\sigma_{\,\mathrm{fit}} \, (\alpha_{\,\lambda})$ includes the formal error inherent to the fitting process, which is due to the error in the detected flux, the removal of broad emission lines in the region $\lambda \geq 121.6 \nm$ -- which makes the wavelength range become patchy --, and the systematic uncertainty due to the decreasing available wavelength range with redshift mentioned above. The systematic effect of the latter in particular can clearly be seen on the overall trend for the uncertainty in the spectral index to increase with redshift.

%

%--------------------------------------------------------------------------------------------------------------
\subsection{Measurement of $\da$} \label{sub:dam}

	For each QSO spectrum, we compute the \emph{total} absorption pixel by pixel in the restframe range $\lambda \in[105, 117] \nm$ according to the expression\footnote{This is just the discrete version of equation (\ref{eq:cfd}).}
	\begin{equation} \label{eq:daz}
		D_{\,\mathrm A}(\zem; \alpha_{\,\lambda}) \equiv 1 - \frac{1}{N_{pix}} \sum_{i=1}^{N_{pix}} \frac{ f_{\,obs}(\lambda_{\,i})}{f_{\,c}(\lambda_{\,i};\alpha_{\,\lambda})} \, ,
	\end{equation}
	where $N_{\,pix}$ is the total number of pixels between \mbox{$\lambda_1 = 105 \cdot (1+\zem) \nm$} and \mbox{$\lambda_2 = 117 \cdot (1+\zem) \nm$}.  The lower limit of this range is chosen as to avoid the wings of the \lyb{} + O{\sc vi} emission lines, while the upper limits is such that the blue wing of the \lya{} emission line is excluded in the computation of $\da$. The measurement of $\da$ for each source together with its corresponding uncertainty are listed in Table \ref{tab:qso} (column five).

	There are two independent sources of uncertainty in the measurement of $\da$: 1) The error in the spectral index inherent to the continuum fitting, \ie{} $\sigma_{\, \mathrm{fit}} \,(\alpha_{\,\lambda})$, and 2) the possibility that the continuum level may have been systematically overestimated due to the change in slope of the power-law at $\lambda = 121.6 \nm$ (cf. Section \ref{sub:con}). We shall refer to the uncertainty in $\da$ due to the first and second sources $\sigma_{\, \mathrm{fit}} \, (\da)$ and  $\sigma_{\, \mathrm{sys}} (\da)$, respectively. The total uncertainty in the measurement of $\da$ --quoted in column five of Table \ref{tab:qso}-- is then assumed to be given by $\sigma^{\,2} \, (\da) = \sigma^{\,2}_{\, \mathrm{fit}} \, (\da) + \sigma^{\,2}_{\, \mathrm{sys}} \, (\da)$.

	As a consistency check, we adopt two different methods to estimate $\sigma_{\, \mathrm{fit}} (\da)$. The first method assumes that the $\pm \sigma$ range for each measurement is given by \mbox{$\da \, [\zem;\alpha_{\,\lambda} \pm \sigma_{\,\mathrm{fit}}\,(\alpha_{\,\lambda})] - \da \, [\zem;\alpha_{\,\lambda}]$}, respectively, for the corresponding values of $ \sigma_{\,\mathrm{fit}}\,(\alpha_{\,\lambda})$ listed in Table \ref{tab:qso}. The second method is based on error propagation, according to which the error $\sigma \, (f)$ in the estimate of a quantity $f\,(x_{\,i})$, which depends on $n$ independent random variables $\{x_{\,i}\}$, each with an uncertainty $\sigma(x_{\,i})$, is given by
	\begin{equation} \label{eq:err}
		\sigma^{\,2}\,(f) = \sum_{\,i=1}^{\,n} \sigma^{\,2}(x_{\,i}) \cdot \left(\frac{\partial f}			{\partial x_{\,i}}\right)^{\,2} \, .
	\end{equation}
	In the case of our measurements of $\da$, only the uncertainty in the index $\alpha_{\,\lambda}$ is relevant, and hence equation (\ref{eq:err}) becomes, with the appropriate notation,
	\begin{equation} \label{eq:err2}
		\sigma_{\, \mathrm{fit}}\, (\da) \equiv \sigma_{\, \mathrm{fit}}\, (\alpha_{\,\lambda}) \cdot \frac{1}{N_{pix}} \sum_{i=1}^{N_{pix}} \, \ln \, (\lambda_{\,i} + \lambda_{\,0}) \, \frac { f_{\,obs}(\lambda_{\,i})}{f_{\,c}(\lambda_{\,i};\alpha_{\,\lambda})} \, ,
	\end{equation}
	where we have used equations (\ref{eq:pow}) and (\ref{eq:daz}). Note that we do not include the term due to the error in the detected flux $f_{\,obs}$, since this is already included in $\sigma_{\, \mathrm{fit}}\, (\alpha_{\,\lambda})$. The error computed according to the first method is asymmetric, as expected, while the error computed using equation (\ref{eq:err2}) approximately (in some cases exactly) corresponds to the arithmetic mean of the former. Since the error estimates computed according to both these methods are consistent with each other, we may use either to estimate the uncertainty in $\da$ due to the uncertainty in the continuum fit; we choose to use the values computed according to method 1, which are listed in Table \ref{tab:qso} (column six). 

	It is not clear based on the results found in the literature (cf. Section \ref{sub:con}) whether the continuum blueward of \lya{} may \emph{always} be overestimated when assuming it to be described by an extrapolation of the power-law redward of \lya{}. However, we will assume that this is the case in order to allow for the possibility that our $\da$ measurements may have been systematically overestimated. In order to compute the corresponding uncertainty in $\da$, we make the following assumptions: First, the continuum in the region blueward of \lya{} shall be described by a power-law as well, but with a smaller spectral index -- \ie{} the continuum is \emph{flatter} -- with respect to the region redward of \lya{}. Since it is not possible to fit a power-law in this regime for the data we use, we assume that both power-laws (above and below \lya{}) are normalised at $\lambda = 121.6 \nm$. Furthermore, and for simplicity, we take the difference in the spectral indices below and above \lya{} to be equal 1 (which corresponds roughly to the findings in the literature quoted in Section \ref{sub:con}). Under these assumptions, the value of $\da$ for our sample turns out to be overestimated on average by less than ten per cent. This value is slightly lower for our measurement of $\da$ at $\zem = 2.71$ than the correction estimated by \citet*{sel03a}. However, they also remark that the exact value of the bias depends on the particular method applied to extrapolate the continuum in the \lya{} Forest region. We hence make the rather conservative assumption that each measurement of $\da$ has an additional uncertainty $\sigma_{\,\mathrm{sys}} \, (\da)$ of ten per cent, and add this uncertainty only to the \emph{low} error bound, since we are assuming that the continuum, and hence $\da$, may have been \emph{over}estimated. We could in principle correct our $\da$ estimates by shifting all measurements by ten per cent towards a lower value, but since it is not clear whether the continuum has been overestimated for \emph{every single} measurement, we prefer to express this bias as an uncertainty. The uncertainty $\sigma_{\, \mathrm{sys}} \, (\da)$ for each measurement is listed in the last column of Table \ref{tab:qso}. 

	Note that we do not correct our measurements for contamination of metal lines. However, this should not introduce a large error, since their contribution is small. For example, \citet{tyt04a} find that they contribute by $2.3\pm0.5$ per cent to the total absorption at $z = 1.9$. The validity of this assumption will be tested in the next section by the comparison of $\da$-measurements to the results of our simulations, in which the absorption due to metal lines is not included.

%--------------------------------------------------------------------------------------------------------------
\section{Results \& Discussion} \label{sec:res}

%--------------------------------------------------------------------------------------------------------------
\subsection{Observations vs. Models}
	
	We compute the evolution of $\da$ in the redshift interval $0.35 < z_{em} < 6.0$ using the models MMC (with and without Lyman limit systems) and BMC presented in the Section \ref{sec:mod}, which include the effect of different populations of absorbers, \ie{} \lya{} forest clouds and Lyman limit absorbers. For comparison, we include also the values of $\da$ computed using \citet{mei06a}'s model (see Section \ref{sec:rev}). We refer to this model as MTC. For the MMC (with and without LLS) and BMC models, we simulate an ensemble of $N_{\,\mathrm{LOS}} = 4 \cdot 10^{\,3}$ \loss{} at fixed redshift, and compute for each of them the flux decrement $\da$ according to equation (\ref{eq:daz}). In this way we get for each given redshift an ensemble of an equal number of $\da$ values for each model, from which we estimate the 50 per cent quantile (median), and the $\pm 34.13, \pm 43.32, \pm 47.72, \pm 49.38$, and $\pm 49.87$ per cent quantiles around the median, which correspond to the $\pm 1, \pm 1.5, \pm 2, \pm 2.5$ and $\pm 3 \, \sigma$ ranges. We do not compute the mean and $\sigma$ ranges in the standard way, since the distribution of $\da$ is strictly speaking unknown a priori. However, according to the results from Section \ref{sec:mod}, the distribution of $\da$ is not too far from a lognormal or even a Gaussian distribution; hence, the identification of mean with median, of $\pm \, \sigma$ range with the $\pm 34.13$ quantile around the median, and so on, is justified. Since the MTC model only provides a (mean) transmission function at each redshift, we compute for this model one single value for $\da$ at each redshift by numerically integrating the corresponding transmission function in the restframe wavelength range $[105,117] \nm$.

\begin{figure}
		\begin{center}
			\resizebox{\hsize}{!}{\includegraphics{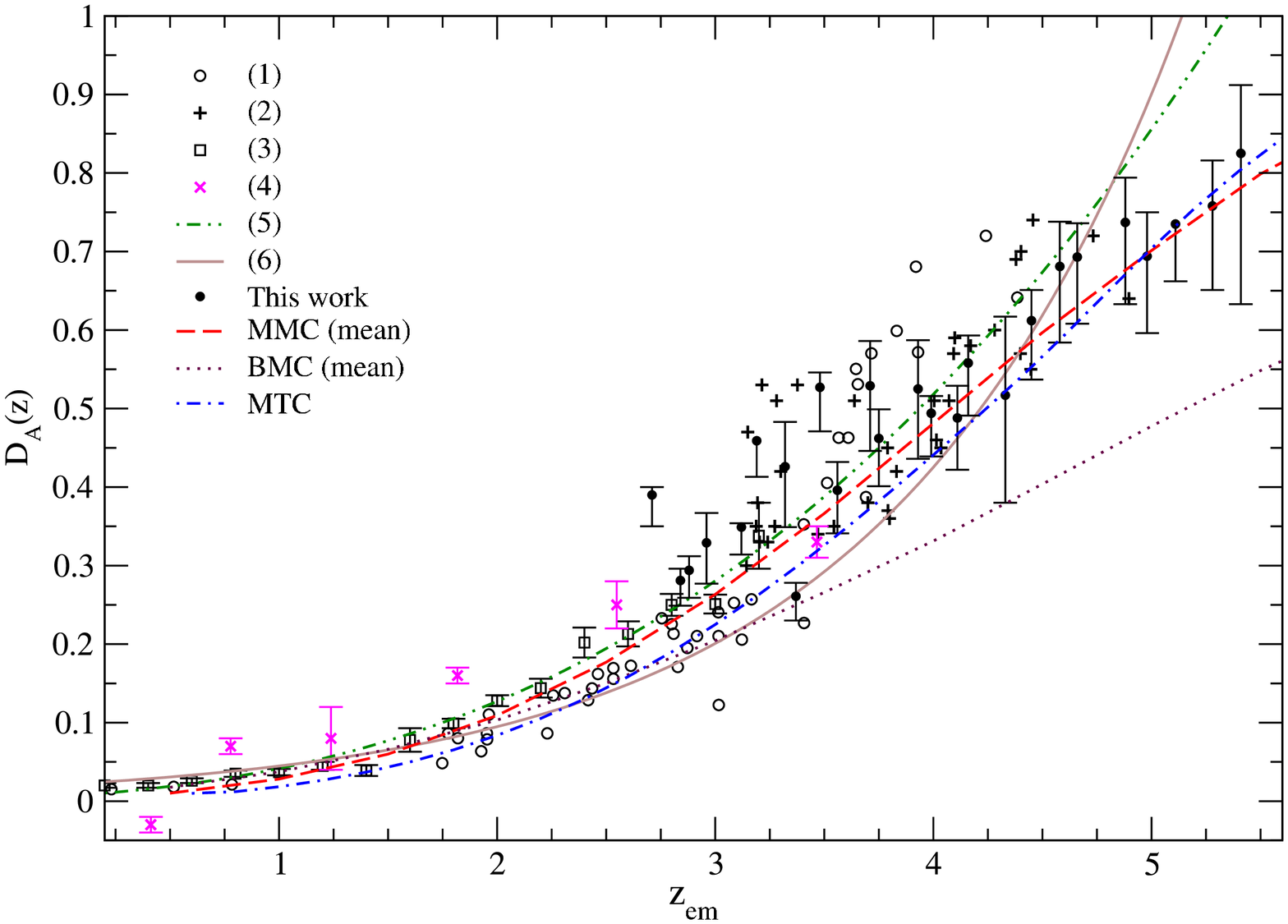}}
			\caption[]{Evolution of the mean $\da$ computed according to the MMC (dashed line), BMC (dotted line), and MTC (dot--dashed line) models, compared to observations performed over the past twenty years by different groups and different methods: (1) \cite*{zuo93c}. (2) \citet*{sch91a}, (3) \citet{kir07a}, (4) \citet*{obr88a}. The data points display $1 \sigma$ error bars. Note that, in spite of the heterogeneity of the approaches to measure $\da$, its mean evolution as computed using the MMC and MTC models matches well the observations over the entire redshift range shown, while they disagree strongly with the BMC model at redshifts $\zem > 3.0$. Below this redshift, the models are practically indistinguishable from each other. For completeness, we include the empirical fits to the evolution of $\da$ from: (5) \citet{zha97a}, and (6) \citet{kir05a}.}
		\label{fig:daz}
		\end{center}
	\end{figure}

	We compare our simulations to measurements of $\da$ accumulated in the literature over the past two decades, performed with different methods and approaches. This compilation is by no means intended to be complete. The reason for choosing these measurements is mainly that they were performed in more or less mutually exclusive redshift ranges which all together cover the range $0 < \zem < 4$. Our measurements extend this redshift range out to $\zem = 5.41$. Therefore, these previous measurements together with our own allow a comparison to the models presented here over a wide redshift range, but with the caveat that using measurements from different groups may introduce an artificial bias in the evolution of $\da$ due to the differences in the particular method used and the different redshift ranges probed in each case, as previously mentioned. The literature data and our measurements are shown together with our model calculations in Figure \ref{fig:daz} (see cited references for details on the corresponding measurements).

	As can be seen in this figure, the predictions for the evolution of the mean $\da$ from all three models, MMC, BMC, and MTC, are practically indistinguishable from each other when compared to observations for redshifts $\zem \lesssim 3$. We do not include in this figure the predictions for $\da$ based on the MMC model without LLSs, since the difference between this and the full MMC model is negligible. At higher redshifts, the values of $\da$ obtained from the MMC and MTC models match the observations quite well. Note that in spite of the fundamentally different approaches and input physics of these models, their results are very similar. As a result, it is not possible yet to discriminate between these models due to the uncertainty and the strong scatter in the observations, especially around $\zem \approx 3.5$ (cf. Section \ref{sec:sca}). In contrast to the MMC and the MTC models, the predictions from the BMC model strongly deviate from the measurements of $\da$ at $\zem > 3$, \emph{underestimating} the absorption.  Since the MMC and BMC models differ only in terms of the input distributions and not in terms of the method, this result suggests that the evolution of the \lya{} absorbers implied by the input distributions of the BMC model is slower than expected from the evolution of $\da$ (cf. equation \ref{eq:pwl}). If used to compute the magnitude changes for high-redshift galaxies due to intergalactic absorption as done by \eg{} \citetalias{ber99a}, the difference in the predicted evolution of $\da$ between the MMC and BMC models corresponds to a difference of slightly more than 0.6 mag in the predicted magnitude change for \eg{} an Sd-type galaxy --with a constant Star Formation Rate-- at $\zem = 4.0$ in the F450W filter. The corresponding difference between the MMC and MTC model amounts to less than 0.3 mag.

	 It is worth mentioning at this point that, in contrast to previous models \citep*[\eg{}][]{zuo93a}, we explicitly avoid normalising in any way the used distributions to match the observed $\da$ at some given redshift, or manipulating the models whatsoever to satisfy any other restriction. We simply take the distributions as reported in the literature, where they were determined directly by means of line statistics by the authors and references therein. We mention this in order to emphasize the rather good agreement between the observations and the evolution of the mean $\da$ predicted by all three models at $\zem < 3$, and by the MMC model at higher redshifts, further reinforced by the similarity between the results of the MTC and the MMC models, which are completely independent from each other.
	\begin{figure}
		\begin{center}
			\resizebox{\hsize}{!}{\includegraphics{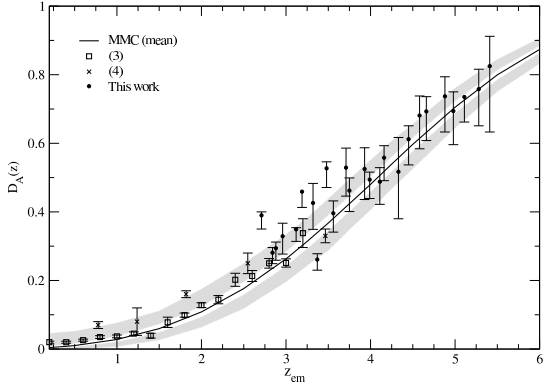}}
			\caption[]{Evolution of the mean $\da$ according to the MMC model (solid line) in the range $0.2 < \zem < 6$. Here we show again our measurements of $\da$ and those from \citet*{obr88a}, together with the most recent measurements from \citet{kir07a}. The data points display $1 \sigma$ error bars. The white and shaded areas around the solid line indicate the intrinsic scatter at the $\pm1$ and $\pm3\,\sigma$ level, respectively, due to variation in the absorption from one random \los{} to another.}
		\label{fig:daz2}
		\end{center}
	\end{figure}

	We include in Figure \ref{fig:daz} the empirical fits of \citet{zha97a} of the form $\da \, (z) = \da^{\,0} \, \mathrm{e}^{\,\alpha \, (1+z)}$ with $\da^{\,0} = 0.01$ and $\alpha = 0.75$ and of \citet{kir05a} of the form $\da \, (z) = A \, (1+z)^{\,\gamma}$ with $A = 0.0062$ and $\gamma = 2.75$. These empirical fits match the observations quite well at $\zem \lesssim 4.5$, and may be useful as a rough estimate of the mean absorption at those redshifts. However, as stated before, they should be taken with caution, especially at high redshifts. While $\da$ asymptotically converges to unity as the redshift approaches the epoch of reionisation, $z_{reion}$, these empirical fits diverge. In contrast, the predicted evolution of $\da$ from all models described above does satisfy the expected and observed asymptotic behaviour. The rate of convergence to this limiting value certainly depends on the particular set of input distributions used, as shown by the comparison between the MMC to the BMC model. In general, the stronger the number density evolution, the faster the convergence. In other words, different input distribution functions imply different values for $z_{\,reion}$, and this information may be used as a further constraint on the accuracy of a particular set of input distributions. Extending the computations of the evolution of $\da$ with the MMC model to redshifts $\zem > 6.0$, it turns out that $\da$ is almost unity at $\zem \approx  7.0$. This value is slightly higher than the value of 6.5 quoted by \citet{fan02a} for the epoch of reionisation.

	The compilation of $\da$ measurements shown in Figure \ref{fig:daz} displays a large scatter at all redshifts, in particular at redshifts $\zem > 3$. Previously, concerns were raised about the possibility that the amplitude of the intrinsic scatter due to cosmic variance may be enhanced by combining measurements performed with different methods and using heterogeneous data samples. Hence, we select a subset from our compilation following the criteria stated below, and compare this observations to the scatter as predicted by our simulations. The results of this comparison are summarised in Figure \ref{fig:daz2}. Here, we show the redshift evolution of $\da$ as computed from the MMC model. We include the 50 per cent quantile, \ie{} the median, and the $\pm 34.13$, and $\pm 49.87$ per cent quantiles, which correspond to the $\pm 1$-, and $\pm 3 \, \sigma$ ranges around the median at each redshift. Again, the difference between the predictions of the MMC with and without LLSs is negligible, and thus we show only the results for the full MMC model. We do not include the BMC model in this analysis in view of the disagreement with the observations at redshifts $\zem > 3$. Our new data subset consists of our own measurements together with the measurements of \citet*{obr88a}, who  use power-law fits to estimate the continuum level of the quasars, and of the measurements from \citet{kir07a}, who fit the continuum locally using b-splines. We select these measurements for the following reasons: \citeauthor*{obr88a}'s data allow a more direct comparison to our data due to the similarity in method to estimate $\da$; \citet{kir07a}'s data are the most recent and accurate measurements of $\da$ available to date. Furthermore, these data sets individually cover the largest redshift ranges, and together they constitute the smallest set of individual measurements fully spanning the redshift range $0.2 < \zem < 5.41$ with some overlap.

	Within the quoted uncertainty, the observations are well matched by the evolution of $\da$ as predicted by the MMC model in the redshift range $0.2 < z_{em} < 5.41$, with the exception of some outliers around $\zem \approx 3$, where the scatter of our observations alone as well as the scatter in the full sample of measurements is large, as can be seen in Figure \ref{fig:daz}. It is apparent from Figure \ref{fig:daz2} that the largest scatter in the observed values of $\da$ at a given redshift cannot be ascribed to the \emph{intrinsic} scatter in the absorption due to variations from one \los{} to another as computed with the MMC model, not even at the $\pm3\,\sigma$ level. Nevertheless, note that this is only the case for \citet*{obr88a}'s and our measurements, while \citet*{kir07a}'s measurements are all nicely contained within the $\pm 3 \, \sigma$ envelope of the model. This suggests that the observed scatter --or at least part of it-- is not real, but only an artifact introduced by combining observations based on different methods to measure $\da$. This assumption is supported by the small amplitude of the scatter in the distribution of high-redshift galaxy colors already found by \citetalias{ber99a} (cf. their Figure 6) for a variety of input distribution functions. Also, the fact that \citet*{kir07a}'s measurements are well matched by the MMC model points to the fact that the particular method chosen by \citet*{obr88a} and ourselves to estimate the continuum in the \lya{} Forest region does introduce a bias in the measurements of $\da$. On the one hand, there is very likely a bias do to the already discussed systematic overestimate of the continuum level. Furthermore, neither \citet*{obr88a} nor we corrected our corresponding measurements for metal-line absorption. These facts together would explain in part why precisely our and their measurements are higher on average than the median value of $\da$ predicted by the MMC model, as can be seen in Figure \ref{fig:daz2}. Even though its effect is expected to be small, including the absorption of metal lines should eventually increase the agreement between the results from the models and the observations.

	There is thus rather strong evidence that most of the amplitude in the observed scatter in $\da$ displayed in Figures \ref{fig:daz} and \ref{fig:daz2} is introduced by combining measurements performed on heterogeneous data samples using different methods. If real, however, the observed scatter would indicate that the models, in particular the MMC model, cannot account for the variation of $\da$ among different \loss{}, even though it reproduces well its \emph{mean} redshift evolution.

	According to the results from our simulations (cf. Section \ref{sec:dda}), it is expected that $\da$ as predicted by the MMC model be lognormally distributed with a high confidence at redshifts $\zem \lesssim 6$, and consequently, that $\ln \da$ and hence the optical depth of \hi{} should obey a Gaussian distribution. However, this result is not supported by the observations given the poor agreement between the scatter predicted by the MMC models and the observed distribution of $\da$ at a given redshift. On the other hand, the expectation that $\da$ should be either lognormally or Gaussian distributed is based solely on the fact that the transmission factor be written in the form of equation (\ref{eq:msk}), which is certainly independent of the particular model accounting for the evolution of the \lya{} absorbers. A different set of input distribution functions may have an effect on the mean value and the spread of the distribution of $\da$ at each redshift, but not the on the form of the distribution. It may be thus possible that a different (more accurate) set of input distribution functions of the absorber properties may help to reconcile the lack of agreement between the \emph{amplitude} of the predicted and the observed scatter, but even in this case the disagreement regarding the \emph{form} of the distribution when compared to previous studies \citep*[\eg{}][]{mad95a,ber03a,tyt04a,mei04a} remains.

	Even though the amplitude of the observed and simulated scatter do not quite match, it is interesting to note that the redshift at which the intrinsic scatter in evolution of $\da$ peaks (cf. Figure \ref{fig:sca}) roughly coincides with the redshift at which the observations show a strong scatter  (cf. Figure \ref{fig:daz}). This redshift matches the redshift at which \citet{ber03a} reported a local decrement in the evolution of the \lya{} optical depth $\tau_{\,\mathrm{eff}}$, recently confirmed by \citet{fau07a}, which has been interpreted by \citeauthor{ber03a} as a signature of the reionisation of \mbox{He {\sc ii}}. Note, however, that at this redshift the $\da$ measurements lie predominantly above the mean value as given by the models (cf. Figure \ref{fig:daz}), and hence the trend suggested by these observations is exactly the opposite from that found by \citet{ber03a}, since a decrease in $\tau_{\,\mathrm{eff}}$ implies a decrease in $\da$. This strongly indicates that the difference in method and the heterogeneity of the samples of the different surveys does introduce an artificial bias in the observed evolution of $\da$, in particular in its scatter.

	We finally want to highlight the following curiosity: By taking a close look at Figures \ref{fig:sca} and \ref{fig:daz2} it becomes apparent that the maximum in the evolution of $\sigma \, (\da)$ roughly coincides with the point of inflection of the curve that describes the expected redshift evolution of the mean $\da$. Mathematically, this would imply that the intrinsic scatter of $\da$ is proportional to the rate of change of the mean $\da$ with redshift, \ie{}
	\begin{equation} \label{eq:der}
		\sigma \, (\da) \propto \frac{\partial \, \langle \da \rangle}{\partial \, z} \, .
	\end{equation}
	Indeed, if one computes numerically the derivative of $\langle \da \rangle$ with respect to $z$, it turns out that it qualitatively matches the evolution of $\sigma\,(\da)$, up to a scale transformation. A physical interpretation may be gained in light of equation (\ref{eq:pwl}): A rapid evolution of the absorbers, quantified by the parameter $\gamma$, implies a stronger evolution of $\da$ with $z$. Since the evolution is however different in general for different \loss{}, this in turn implies a larger variation in the absorption from \los{} to \los{}, and thus a larger value of $\sigma\,(\da)$. This is mathematically consistent with the fact that the scatter given by the derivative of equation (\ref{eq:pwl}) with respect to $z$ is proportional to $\gamma$.

%______________________________________________________________
\section{Summary \& Conclusions} \label{sec:con}

\begin{enumerate}
     \item We measured the cosmic flux decrement $\da$ and its uncertainty due to statistical and systematic errors for 25 QSOs of the SDSS DR5 catalog in the redshift range $2.71 \leq \zem \leq 5.41$.
     \item We modeled the redshift evolution of the mean $\da$ and its distribution at each given redshift in a Monte Carlo fashion, adopting two among the various models presented by \citet*{ber99a}. We found that the predictions of the MMC model for the evolution of the mean $\da$ reproduce well the observations in the range $0.2 < \zem < 5.41$, in contrast to the BMC model. We conclude from this that the underlying input distributions of the BMC model may not accurately account for the evolution of the \lya{} absorption on the IGM. Hence, estimates of the impact of the intergalactic attenuation on the photometric properties of high-redshift galaxies using this particular model should be taken with caution. Incidentally, by showing the rather good agreement between one particular set of simulations and the data, we show the power of the relatively simple approach used here to model the effect of intergalactic absorption, as compared to models based on hydrodynamical simulations which are by far more complex.
      \item Through the comparison between the MMC and BMC models, we showed that different input distribution functions may have very different predictions regarding the evolution of $\da$, a quantity which is intimately related --and hence very sensitive-- to the amount of \hi{} present along the \los{}. Therefore, we state that any model of the intergalactic absorption should first of all reproduce the observed evolution of $\da$ and its distribution before it is used to \eg{} correct synthetic or observed spectra for intergalactic absorption.
     \item The results from our simulations suggest that the distribution of $\da$ at a given redshift be well described by a lognormal distribution at low redshifts and by a Gaussian distribution at high redshifts, in agreement with our theoretical expectation based on the fact that the absorption is mathematically expressed as the product of small, statistically independent factors. This result implies that at redshifts where $\da$ is distributed lognormally, the effective optical depth of the intergalactic \hi{} should obey a normal distribution, contrary to the results of previous studies. However, this result should be taken with caution in light of the fact that the models, in particular the MMC model, cannot reproduce the amplitude of the observed scatter in $\da$.
     \item We argue that most of the observed scatter in $\da$ is introduced by combining measurements based on different methods. Nevertheless, the reason for the lack of agreement between the scatter predicted by the MMC model and the observed distribution of $\da$ is not ultimately settled. Because of this and until this discrepancy is clarified, we may warn about using any attenuation model based on the input distributions of the MMC model to estimate magnitude changes in the spectra of background sources.
     \item A larger, homogeneous sample of accurate measurements of $\da$ over a wide redshift range is needed in order to allow for a more faithful comparison to models, and in particular, to determine the intrinsic \emph{form} and \emph{amplitude} of the distribution of $\da$ as a function of redshift.
\end{enumerate}

%______________________________________________________________
\section*{Acknowledgments}
      
      We want to thank our referee M. Bershady for his insightful comments and suggestions which helped improving our work. We are also grateful to Avery Meiksin for computing a special set of tables of his transmission functions, and to Nico Bissantz for useful suggestions about the statistical treatment of the data. TTG acknowledges financial support from the \emph{Mexican Council for Science and Technology} (CONACyT), and the \emph{Deutsche Forschungsgemeinschaft} (DFG) through Grant DFG-GZ: Ri 1124/5-1.\\
      
      Funding for the SDSS and SDSS-II has been provided by the Alfred P. Sloan Foundation, the Participating Institutions, the National Science Foundation, the U.S. Department of Energy, the National Aeronautics and Space Administration, the Japanese Monbukagakusho, the Max Planck Society, and the Higher Education Funding Council for England\footnote{The SDSS Web Site is http://www.sdss.org/.}.\\

	The SDSS is managed by the Astrophysical Research Consortium for the participating Institutions. The Participating Institutions are the American Museum of Natural History, Astrophysical Institute Potsdam, University of Basel, University of Cambridge, Case Western Reserve University, University of Chicago, Drexel University, Fermilab, the Institute for Advanced Study, the Japan Participation Group, Johns Hopkins University, the Joint Institute for Nuclear Astrophysics, the Kavli Institute for Particle Astrophysics and Cosmology, the Korean Scientist Group, the Chinese Academy of Sciences (LAMOST), Los Alamos National Laboratory, the Max-Planck-Institute for Astronomy (MPIA), the Max-Planck-Institute for Astrophysics (MPA), New Mexico State University, Ohio State University, University of Pittsburgh, University of Portsmouth, Princeton University, the United States Naval Observatory, and the University of Washington.\\

%__________________________________________________________________

%--------------------------------DOCUMENT END-----------------------------------
\label{lastpage}
\end{document}